\documentclass[12pt]{article}
\usepackage{amsthm,amsmath,latexsym,amssymb,amsfonts,amscd}
\usepackage{graphics,lscape,fancyhdr,array,stmaryrd,euscript,wrapfig}
\usepackage{empheq}
\usepackage{verbatim,slashed}
\numberwithin{equation}{section}
\usepackage{hyperref,setspace,tikz}
\usepackage{tikz-feynman}
\tikzfeynmanset{compat=1.1.0}
\usepackage[numbers,sort&compress]{natbib}
\usepackage[nottoc]{tocbibind}
\usepackage{cancel}

\newcommand{\pl}{\partial}
\newcommand{\besubeqs}{\begin{subequations}}
\newcommand{\esubeqs}{\end{subequations}}

\newcommand{\aA}{{\ensuremath{A}}}
\newcommand{\aB}{{\ensuremath{B}}}
\newcommand{\aC}{{\ensuremath{C}}}

\newcommand{\brk}{{{\bar{k}}}}

\newcommand{\fA}{\mathbf{f}}
\newcommand{\PP}{{\mathbb{P}}}

\newcommand{\fud}[2]{{}^{#1}{}_{#2}\,}
\newcommand{\fdu}[2]{{}_{#1}{}^{#2}\,}
\newcommand{\fudu}[3]{{}^{#1}{}_{#2}{}^{#3}\,}

\newcommand{\PPb}{{\bar{\mathbb{P}}}}

\newcommand{\Tr}{{\mathrm{Tr}\,}}

\newcommand{\bri}{{\bar{i}}}
\newcommand{\brj}{{\bar{j}}}
\newcommand{\brA}{{\bar{1}}}
\newcommand{\brB}{{\bar{2}}}
\newcommand{\brC}{{\bar{3}}}
\newcommand{\brD}{{\bar{4}}}

\DeclareMathOperator{\sign}{sg}
\newcommand{\Cg}{{\mathcal{C}}}
\newcommand{\Bg}{{\mathcal{B}}}

\newcommand{\ladder}{
\begin{tikzpicture}[line width=0.8pt, scale=0.5]
\draw (0,0) -- (0,2);
\node[above]  at (0,2) {$a_{\sigma_2}$};
\node[left]  at (-2,0) {$a_1$};
\draw (-2,0) -- (2.5,0);
\draw (2,0) -- (2,2);
\node[above]  at (2,2) {$a_{\sigma_3}$};
\draw[dashed] (2.5,0) -- (3.5,0);
\draw (3.5,0) -- (8,0);
\draw (4,0) -- (4,2);
\node[above]  at (4,2) {$a_{\sigma_{n-2}}$};
\draw (6,0) -- (6,2);
\node[above]  at (6,2) {$a_{\sigma_{n-1}}$};
\node[right]  at (8,0) {$a_{n}$};
\end{tikzpicture}
}

\begin{document}
\hfill
\vskip 0.01\textheight
\begin{center}
{\large\bfseries 
Amplitudes in self-dual (higher-spin) theories}

\vspace{0.4cm}

\vskip 0.03\textheight
\renewcommand{\thefootnote}{\fnsymbol{footnote}}
Mattia Serrani \& Evgeny \textsc{Skvortsov}

\vskip 0.03\textheight

{\em Service de Physique de l'Univers, Champs et Gravitation, \\ Universit\'e de Mons, 20 place du Parc, 7000 Mons, 
Belgium}\\
\vspace*{5pt}

\renewcommand{\thefootnote}{\arabic{footnote}}
\end{center}

\vskip 0.02\textheight

\begin{abstract}
Self-dual theories are powerful toy models of their completions. It was shown recently that there are infinitely many SD-theories once massless higher-spin fields are allowed. The maximal SD-theory is chiral higher-spin gravity. Following the recent [2602.12176] we show that all SD-theories, including those with massless higher-spin fields, have nontrivial tree-level amplitudes in Kleinian signature or complex Minkowski kinematics. The results are obtained both in the light-cone and spinor-helicity approaches. Within celestial holography, the nontriviality of amplitudes in chiral higher-spin gravity provides the missing ingredient needed to complete the celestial analogue of the vector-model/higher-spin AdS/CFT duality. 
\end{abstract}

\newpage
\tableofcontents
\newpage
\section{\label{sec:intro}Introduction}
Self-dual theories have been useful approximations and toy models of their completions for many years, most notably within the twistor approach. One remarkable property of the self-dual truncation is that it is not based on any kind of weak field/small coupling expansion. Recently, it has been shown \cite{Ponomarev:2016lrm, Ponomarev:2017nrr, Krasnov:2021nsq,Monteiro:2022xwq,Ponomarev:2024jyg,Serrani:2025owx} that the landscape of self-dual theories is considerably enlarged if one allows for higher-spin fields, which leads to self-dual/chiral higher-spin gravities. 

The idea of higher-spin gravity (HiSGRA) was developed to shed light on the quantum gravity problem, where one can argue that extended symmetries should compete with and eventually win over counterterms/UV-divergences: a large enough symmetry should forbid all possible UV divergences \cite{Fronsdal:1978rb,Fradkin:1986ka}. In addition, since the first obstacles to quantizing gravity arise in the UV regime, one can argue that all states can effectively be treated as massless.\footnote{Of course, string theory is a counterexample since the mass spectrum is unbounded from above. } 

Therefore, without abandoning the concept of particles, one can look for theories with massless fields that contain gravity, which itself is a massless spin-two state. In fact, $s=2$ is a threshold in the sense that having at least one massless higher-spin state, $s>2$, usually requires all spins to be present, see e.g. \cite{Flato:1978qz,Berends:1984wp,Fradkin:1986ka,Maldacena:2011jn,Boulanger:2013zza}.\footnote{There are notable exceptions in $3d$, where massless higher-spin fields are topological,  \cite{Blencowe:1988gj,Bergshoeff:1989ns,Campoleoni:2010zq,Henneaux:2010xg,Grigoriev:2020lzu,Pope:1989vj,Fradkin:1989xt,Grigoriev:2019xmp} and, recently, in $4d$ \cite{Serrani:2025owx} and even in higher even dimensions \cite{Basile:2024raj}.  } However, upon closer examination, nontrivial interactions involving massless higher-spin fields are often in conflict with fundamental field-theoretic principles, such as unitarity and locality. These tensions may indicate that any ``healthy'' quantum gravity model does not quite obey field theory rules and should be nonlocal. The potential benefits of HiSGRA are substantial since the higher-spin symmetry should be powerful enough to render everything UV-finite and there are indications this might be true, see e.g. \cite{Skvortsov:2020wtf,Skvortsov:2018jea,Skvortsov:2020gpn,Tsulaia:2022csz,Ponomarev:2022atv}.  

While direct attempts to construct a higher-spin gravity are always obstructed by non-locality of a sort,\footnote{This is definitely true when one has propagating massless higher-spin fields, see e.g. \cite{Bekaert:2010hp,Roiban:2017iqg,Ponomarev:2017nrr,Serrani:2026dbs,Dempster:2012vw,Bekaert:2015tva,Maldacena:2015iua,Sleight:2017pcz,Ponomarev:2017qab} for a few out of many no-go's. The situation in flat and constant curvature spacetimes is exactly the same, but it is somewhat easier to see where the problem is coming from in AdS thanks to holography \cite{Bekaert:2015tva,Maldacena:2015iua,Sleight:2017pcz,Ponomarev:2017qab}. However, by sacrificing propagating degrees of freedom, one can have nontrivial models in $3d$ and $2d$ with \cite{Sharapov:2024euk,Bekaert:2025azj} and without \cite{Blencowe:1988gj,Bergshoeff:1989ns,Campoleoni:2010zq,Henneaux:2010xg,Grigoriev:2020lzu,Pope:1989vj,Fradkin:1989xt,Grigoriev:2019xmp,Alkalaev:2013fsa,Alkalaev:2020kut,Alkalaev:2019xuv} low-spin propagating degrees of freedom.} a recently discovered ``loophole'' is that there exists chiral higher-spin gravity \cite{Metsaev:1991mt,Metsaev:1991nb,Ponomarev:2016lrm} as well as higher-spin extensions of self-dual Yang-Mills theory (SDYM) and self-dual gravity (SDGR) \cite{Ponomarev:2017nrr,Krasnov:2021nsq}. The complete landscape of such theories is yet to be described \cite{Ponomarev:2017nrr,Monteiro:2022xwq,Serrani:2025owx}. What is clear is that all of the above theories are self-dual ones of some kind \cite{Ponomarev:2017nrr,Serrani:2025oaw}.   

Self-dual theories can be characterized by several useful properties: (a) tree-level amplitudes vanish in flat space for generic kinematics; (b) loop amplitudes are simple and UV-finite; (c) the interactions are confined to cubic terms (in flat space) and all vertices have a definite sign of the total helicity, say positive $\lambda_1+\lambda_2+\lambda_3>0$, e.g. SDYM and SDGR have only vertices of type $(+s,+s,-s)$, $s=1,2$; (d) self-dual theories have a formulation that justifies the word ``self-dual''; (e) they admit natural formulations on twistor space; (f) they are integrable; (g) solutions of self-dual theories are also solutions of their completions. 

Recently, two other characteristics of self-dual (SD) theories have been found: (h) they satisfy the celestial OPE associativity constraint \cite{Ren:2022sws,Serrani:2025oaw}; (i) behind any self-dual theory there is a certain ``kinematic algebra'', which is a Lie algebra completely describing the theory \cite{Ponomarev:2017nrr}, see also \cite{Monteiro:2011pc,Serrani:2025oaw}. In the light-cone gauge, where SD-theories are especially simple, the kinematic algebra is obtained by ``dividing'' the cubic vertices by the SDYM vertex. Therefore, all SD-theories in the light-cone gauge look like SDYM times some kinematic algebra.

\begin{figure}[h]
    \centering
    \includegraphics[width=0.7\linewidth]{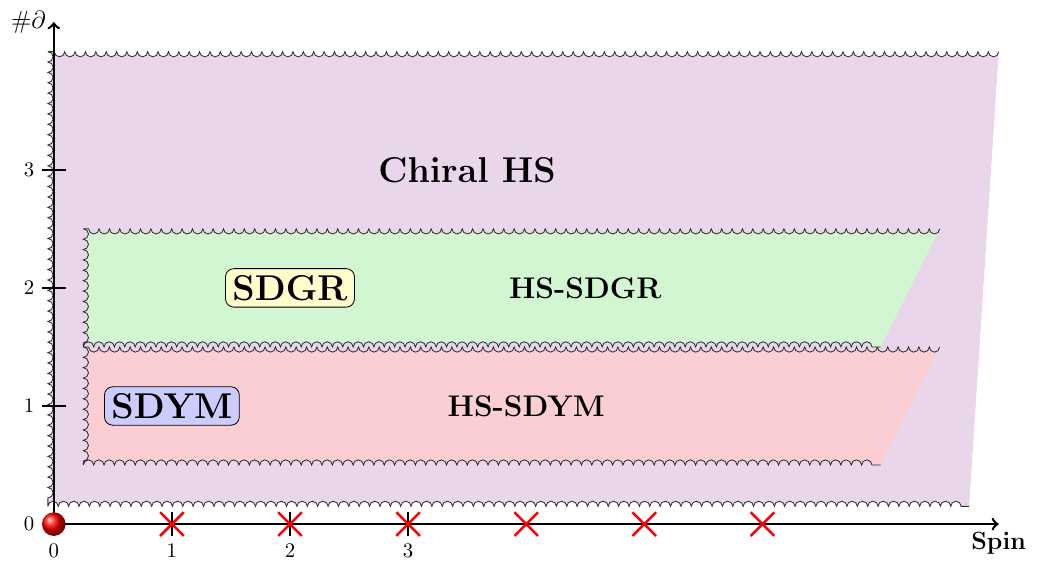}
    \caption{The map of SD-theories with a few basic ones: SDYM, HS-SDYM, SDGR, HS-SDGR and Chiral HiSGRA. Infinitely many more classes can be found in \cite{Ponomarev:2017nrr,Serrani:2025owx}.}
    \label{fig:map}
\end{figure}

While low-spin SD-theories are covered by SDYM, SDGR and supersymmetric extensions thereof, it turned out that there is a ``zoo'' of higher-spin gravities that are ``as good as'' all the other SD-theories, see Fig. \ref{fig:map} for a few of them.\footnote{Historically, it was noted in \cite{Devchand:1996gv} that SD-theories do not restrict the number of SUSYs, which gives some HS-extensions of SDYM and SDGR. } There exists the maximal SD-theory, chiral higher-spin gravity, that has all $\lambda_1+\lambda_2+\lambda_3>0$ interactions and all spins/helicities present in the spectrum. If one seeking a generic SD theory by starting with some putative spectrum/interactions, then most likely the unique Lorentz-invariant completion will be the Chiral HiSGRA. However, there are still infinitely many SD-theories that have higher spins and finitely many fields. 

All members of the ``SD-zoo'' share the main characteristic properties of SD-theories described above. Nevertheless, it is Chiral HiSGRA that has good chances to be embedded into various (A)dS/CFT and flat/CFT dualities \cite{Sharapov:2022awp,Aharony:2024nqs,Jain:2024bza}. It is an ``M-theory'' of all SD-ones in the sense that it contains all spins/helicities and all interactions compatible with self-duality. All other theories can be found inside it, e.g. as contractions \cite{Ponomarev:2017nrr}. 

The main obstacle for flat space applications of HiSGRA has always been the triviality of the flat space S-matrix, which is essentially a consequence of Weinberg's low energy theorem. Possible ways out would be to consider form-factors or nontrivial backgrounds, e.g. a black hole. The simplest examples of nontrivial backgrounds are (anti)-de Sitter spaces, where SD-theories have nontrivial amplitudes \cite{Armstrong:2020woi, Chowdhury:2024dcy,Skvortsov:2026gtq}. A more direct way out follows from the recent \cite{Guevara:2026qzd}. It was shown in \cite{Guevara:2026qzd} that YM has nontrivial $(++...+-)$-amplitudes, i.e. all-plus one-minus, in the Kleinian signature or complex Minkowski kinematics, see also \cite{Guevara:2026qwa} for SDGR. These amplitudes are shared by SDYM and YM.  

This powerful observation has an immediate consequence for all SD-theories at once --- all SD-theories do have nontrivial tree-level amplitudes! This includes numerous SD-theories with higher-spin fields. We prove this by reformulating the double-copy construction in terms of the DDM-ordering \cite{DelDuca:1999iql,DelDuca:1999rs}. The DDM-ordering is the one that operates with structure constants $f^a_{bc}$ of a Lie algebra without having to assume a $\Tr([T_b,T_c] T_a)$-representation. One simple direct consequence is that all SD-theories, e.g. SDGR, have amplitudes that are directly expressible in terms of those of SDYM. Various other consequences of this observation are discussed, including celestial/flat holography. We obtain these results both in the light-cone gauge and in spinor-helicity formalism.

This paper is organized as follows. In section \ref{sec:SDYM}, after low-point examples of collinear amplitudes in SDYM \cite{Guevara:2026qzd}, we prove that HS-SDYM has the same amplitudes dressed with a simple kinematic factor. Self-dual theories, including lower- and higher-spin examples, are reviewed in Section \ref{sec:SDall}, where we also recall the notion of the kinematic algebra \cite{Monteiro:2011pc,Ponomarev:2017nrr,Monteiro:2022lwm,Monteiro:2022xwq,Serrani:2025oaw}. With the help of the DDM-ordering we show that the amplitudes and Berends--Giele currents for all SD-theories can be expressed in terms of those of SDYM, which is demonstrated in Section \ref{sec:BG}, where we also give some convincing examples. In Section \ref{sec:spinorhell} we obtain an equivalent but manifestly Lorenz-invariant result by using the CSW off-shell extension of the spinor-helicity on-shell vertices. Discussion and conclusions can be found in Section \ref{sec:conclusions}, where we speculate how the obtained results allow one to build a celestial analogue of the higher-spin/free vector model duality \cite{Klebanov:2002ja,Sezgin:2002rt,Leigh:2003gk} following \cite{Ponomarev:2022atv,Ponomarev:2022qkx,Ponomarev:2022ryp}. There are a couple of technical appendices.

\section{Amplitudes in (HS-)SDYM}
\label{sec:SDYM}
The amplitudes studied in \cite{Guevara:2026qzd} are of type $(+...+-)$ and are shared by YM and SDYM. Therefore, let us confine ourselves to SDYM. The Chalmers--Siegel action \cite{Chalmers:1996rq} is\footnote{We use the two-component spinor language: $A,B,...=1,2$, $A',B',...=1,2$, the indices are raised/lowered by $\epsilon^{AB}=-\epsilon^{BA}$ and $\epsilon^{A'B'}$ as $\xi^A=\epsilon^{AB}\xi_B$, $\xi^A\epsilon_{AB}=\xi_B$. If a spin-tensor is symmetric or needs to be symmetrized in some indices, the indices can be denoted by the same letter, e.g. $F^{AA}\equiv F^{A_1 A_2}=F^{A_2 A_1}$.  } 
\begin{align}
   S&=\int \Psi^{AA}F_{AA} -\tfrac{\epsilon}{2}\Psi_{AA} \Psi^{AA}\,.
\end{align} 
It depends on two fields: a gauge field $\Phi_{AA'}\sim A_\mu$ and $\Psi^{AB}=\Psi^{BA}$, which is often realized as a self-dual two-form. Here, $F_{AA}=\partial_{AC'} \Phi\fdu{A}{C'}+i \Phi_{AC'}\Phi\fdu{A}{C'}$ is $F_+$ in the two-component spinor language, which is the most convenient one to deal with self-dual theories. 
For $\epsilon\neq0$ the field $\Psi^{AA}$ is an auxiliary field that can be solved for from $F_{AA}=\epsilon\Psi_{AA}$ to find $(F_{AA})^2$ for the Lagrangian. The latter is the YM Lagrangian up to a total derivative. For $\epsilon=0$ we have SDYM, where $\Psi_{AA}$ is not a Lagrange multiplier anymore, but will carry one degree of freedom, whereas $\Phi^{AA'}$ carries the other one. SDYM has the same number of degrees of freedom as YM. The SDYM action has a very simple higher-spin extension \cite{Krasnov:2021nsq}, which we will discuss later. In the Chalmers--Siegel form the equations of motion for $\Phi^{AA'}$ read
\begin{align}\label{SDYMeom}
    p\fud{A}{C'}\Phi^{AC'}&= \Phi^{AC'} \Phi\fud{A}{C'}  && \Longrightarrow && \Phi^{AA'}= (p^2)^{-1}{p\fdu{A}{A'} (\Phi^{AC'} \Phi\fud{A}{C'})}\,.
\end{align}
The equations are the momentum space version of the self-duality constraint $F_{AA}=0$. We kept the color indices implicit and displayed the naive propagator $\slashed p/p^2$. The equations can be solved by iterations, which is essentially the Berends--Giele method \cite{Berends:1987me}. 

\subsection{Warm ups}
\label{subsec:warmup}
It is very instructive to compute the first few amplitudes to see where the collinear kinematics is coming from.  We fix the plane waves as $\Phi^{AA'}_i={r^A \bar{i}^{A'}}/{(ri)}$, where $r^A$ is a reference spinor. The $n$-point amplitude is defined to be just the contraction $\Psi_{AA}^n(\Phi^{AC'} \Phi\fud{A}{C'})_{12...n-1}$ of the r.h.s.  of \eqref{SDYMeom} evaluated to order $(n-1)$ with the plane wave $\Psi_{AA}^n=n_An_A$ of the $n$-th gluon. For the three-point, we directly get\footnote{As different from the $[ij]$, $\langle ij\rangle$ notation, we choose to always write $(ij)\equiv i^C j_C$ and indicate the type of spinors by an overbar, if needed. This is also justified by (A)dS/CFT applications where the symmetry is broken down to the boundary Lorentz algebra, which makes primed and unprimed spinors take values in isomorphic representations. }
\begin{align}
    \mathcal{A}_3= \frac{(r3)^2 (\brA \brB)}{(r1)(r2)}\,.
\end{align}
The on-shellness implies $(ij)(\bri \brj)=0$, which can be made more manifest: every amplitude is followed by the delta-function of the total momentum. This measure can be rewritten as
\begin{align}
    \delta^4\left(\sum i_A \bri_{A'}\right) &= \delta^2\left(\sum (ri) \bri_{A'}\right)\delta(13)\delta(23)\frac{(r3)^2}{|(\brA \brB)|}\,,
\end{align}
where we used $\delta^2(a\lambda_A+ b\mu_A)= \delta(a)\delta(b)/|(\lambda \mu)|$. Therefore, the complete three-point amplitude, including the momentum conservation, reads ($\sign(x)$ is a shorthand for the sign-function)
\begin{align}\label{cubicAmp}
    \mathcal{A}_3=\frac{(r3)^4}{(r1)(r2)} \sign(\brA\brB) \delta(13)\delta(23)\,\delta^{0|2}\,,
\end{align}
where the leftover from the momentum conservation is 
\begin{align}
    \delta^{0|2}=\delta^2\left(\sum_i (ri) \bri_{A'}\right)\,.
\end{align} 
At the next order, we solve for $\Phi_{ij}$ that is sourced by $\Phi_i$ and $\Phi_j$
\begin{align}\label{threepoint}
    \Phi_{ij}^{AA'}&= \frac{1}{p_{ij}^2+i\epsilon}\frac{r^C p\fudu{ij}{C}{A'}}{(ri)(rj)} r^A (\bri \brj)\,.
\end{align}
This is to be used to get the quartic amplitude
\begin{align}
    \mathcal{A}_4&= 4_A4_A \left(\Phi_{12}{}^{AC'} \Phi_3\fud{A}{C'}+ \Phi_{1}{}^{AC'} \Phi_{23}\fud{A}{C'}\right)\,.
\end{align}
Let us consider the first term, the s-channel:
\begin{align}
   \mathcal{A}_s&= \frac{(r4)^2}{(r1)(r2)(r3)} \frac{-r^C p^{12}_{CC'} \brC^{C'}}{p_{12}^2+i\epsilon} (\brA \brB) \,.
\end{align}
The pole part, $\mathrm{PV}(1/p^2)$, when combined with the t-channel, gives $0$ on-shell (after applying a Fierz identity once), as is well-known. Let us consider the next term by replacing $1/(x+i\epsilon)$ with $\delta(x)$ (to keep the expressions simple and free of $i\pi$-factors):
\begin{align}
    -\frac{(r4)^2}{(r1)(r2)(r3)} (r^C p^{12}_{CC'} \brC^{C'}) \sign(\brA \brB) \delta (12)\, \delta^{0|2}\,\delta^2\left(\sum (\mu i) \bri_{A'}\right) (r \mu)^2\,.
\end{align}
Here, the conservation was split into two parts, along $r$ and $\mu$ directions. We choose $\mu=1$ to get $\delta^2((12)\brB_{A'}+(13)\brC_{A'} +(14) \brD_{A'})$, where we can use $\delta(12)$ immediately to get
\begin{align}
    -\frac{(r4)^2}{(r1)(r2)(r3)} (r^C p^{12}_{CC'} \brC^{C'}) \sign(\brA \brB) \delta (12) \delta(13)\delta (14) \frac{(r 1)^2}{|(\brC \brD)|} \,\delta^{0|2}\,.
\end{align}
Now, the deltas imply that all $1234$ are parallel. Let us massage them into $\delta (14)\delta (24)\delta (34)$, which gives Jacobian $(r4)^2/(r1)^2$.  We also replace $p_{12}=-p_{34}$ and, hence, $(r^C p^{12}_{CC'} \brC^{C'})=-(r4)(\brC \brD)$. The final form of the s-channel reads \cite{Guevara:2026qzd}
\begin{align}
    \frac{(r4)^5}{(r1)(r2)(r3)} \sign(\brA \brB)\sign(\brC \brD) \,\delta (14)\delta (24)\delta (34)\, \delta^{0|2}\,.
\end{align}
More generally, one can consider different Green's functions and amplitudes associated with them. Any two Green's functions differ by a homogeneous solution. For example, one can choose $G_0=\mathrm{PV}\frac{1}{p^2}$ as a reference Green's function. Then, any other Green's function $G_\alpha$
\begin{align}
    G_\alpha = G_0(p)-i \pi\delta(p^2)\alpha(p)=\mathrm{PV}\frac{1}{p^2}-i \pi\delta(p^2)\alpha(p)
\end{align}
is parameterized by a function $\alpha(p)$ defined on the shell $\delta(p^2)$, i.e. $\alpha(p)=\alpha(p_0,|\vec{p}|)$. For example, $\alpha=1$ for Feynman's propagator and $\alpha=\pm \sign(p^0)$ for advanced/retarded propagators.\footnote{The Lorentz invariant options stop at $c_1+ c_2 \sign(p_0)$. However, the thermal vacuum violates Lorentz symmetry and depends on an additional vector. } Finally, with the most general boundary conditions, the essential part of $\mathcal{A}_s$ is  
\begin{align}
    \mathcal{A}_s&=\frac{(r4)^5}{(r1)(r2)(r3)} \sign(\brA \brB)\alpha(p_{12})\sign(\brC \brD) \,\delta (14)\delta (24)\delta (34)\, \delta^{0|2}\,.
\end{align}
Likewise, for the t-channel we get
\begin{align}
    \mathcal{A}_t&=\frac{(r4)^5}{(r1)(r2)(r3)}  \sign(\brB \brC)\alpha(p_{23})\sign(\brD \brA) \,\delta (14)\delta (24)\delta (34)\, \delta^{0|2}\,.
\end{align}
We note that one $\sign$-function comes from the measure and another one comes from the cut-propagator. In general, for an $n$-point amplitude $(n-3)$ $\sign$-functions come from cut-propagators and should be accompanied by $\alpha(\bullet)$ for a generic Green's function $G_\alpha$.\footnote{There are additional contributions that do not correspond to maximal cuts (maximal cuts are just the cubic graphs with all internal lines replaced by $\delta(p^2)$.} 

Let us make one comment that will be useful for a higher-spin generalization. At the order $n$, the Berends--Giele current is (here, $p\equiv p_{1...n}=p_1+...+p_n$)
\begin{align}
    \Phi^{AA'}_{1...n}&= (p^2)^{-1}{p\fdu{A}{A'} \sum_{i}(\Phi^{AC'}_{1...i}\, \Phi_{i+1...n}^{\vphantom{A}}{}\fud{A}{C'})}\,.
\end{align}
An observation is that for the choice of the polarization tensors we made, $\Phi^{AA'}$ is always of the form $r^A\Phi^{A'}$, cf. \eqref{threepoint}. Therefore, effectively, one needs to iterate
\begin{align}\label{SDYMBG}
    \Phi^{A'}_{1...n}&= (p^2)^{-1}{r^A p\fdu{A}{A'} \sum_{i}(\Phi^{C'}_{1...i}\, \Phi^{\phantom{A}}_{i+1...n}{}_{C'})}\,,
\end{align}
with the initial polarization of the form $\Phi^{A'}_i={\bar{i}^{A'}}/{(ri)}$. This is an ``effective field theory'' that corresponds to the Berends--Giele recursion with the particular choice of plane waves that make $r^A$ factor out.\footnote{One can write down an ``effective action'' $\int \Psi^{A}(\pl_{AA'} \Phi^{A'} +r_A \Phi^{C'}\Phi_{C'} )$. } We see that ``effectively'' SDYM is a theory of $\Phi^{A'}$ rather than of $\Phi^{AA'}$.

\subsection{HS-SDYM}
\label{sec:HSSDYM}
HS-SDYM is a straightforward higher-spin extension of SDYM \cite{Ponomarev:2017nrr,Krasnov:2021nsq}. Let us start with free higher-spin fields. The equations of motion for massless (higher-spin) $s\geq1$ fields, which originate from twistor theory \cite{Penrose:1965am,Hughston:1979tq,Eastwood:1981jy,Woodhouse:1985id}, can be written as 
\begin{align}\label{firstphi}
\begin{aligned}
        \partial\fud{A}{C'} \Phi^{A(2s-1),C'}&=0\,,\qquad\qquad && \delta \Phi^{A(2s-1),A'}=\partial^{AA'}\xi^{A(2s-2)}\,, \\
    \pl\fdu{B}{A'} \Psi^{BA(2s-1)}&=0 \,.
\end{aligned}
\end{align}
Here, $A(k)\equiv A_1...A_k$ denotes a group of $k$ symmetric indices. 
One, say, positive helicity is described by a gauge field $\Phi^{A(2s-1),C'}$, and the opposite helicity is taken care of by $\Psi^{A(2s)}$. The polarization vectors of (SD)YM can be extended to any $s$:
\begin{align}\label{polarizPhi}
    \epsilon^+_{A(2s-1),A'}(k) &=  \frac{r_{A_1} \cdots r_{A_{2s-1}} \bar k_{A'}}{(rk)^{2s-1}} \, , && \epsilon^{A(2s)}_{-}(k) = k^{A_1} \ldots k^{A_{2s}}\, .
\end{align}
The action is more compactly written with the help of differential forms and generating functions. The free action for $s\geq1$ is \cite{Krasnov:2021nsq}
\begin{align}
    S&= \int \Psi^{A(2s)} H_{AA} \wedge d\omega_{A(2s-2)}\,.
\end{align}
Here, $H^{AB}=e\fud{A}{C'}\wedge e^{BC'}$ is the basis of self-dual two-forms built from the vierbein one-form $e^{AA'}$. The gauge field $\Phi^{A(2s-1),A'}$ is embedded into a one-form $\omega_{A(2s-2)}$. This form has one extra irreducible component as compared to $\Phi^{A(2s-1),A'}$:
\begin{align}
\omega^{A(2s-2)}\equiv e_{BB'}\Phi^{A(2s-2)B,B'}+e\fud{A}{B'}\Theta^{A(2s-3),B'}\,.
\label{omPhiTheta}
\end{align}
The second component can be removed thanks to the additional symmetry the action has
\begin{equation}
    \delta \omega^{A(2s-2)}= d \xi^{A(2s-2)} +e\fud{A}{C'} \eta^{A(2s-3),C'}\,.
\end{equation}
It is convenient to introduce a generating function of all (higher-spin) fields
\begin{align}
    \omega(y)&= \sum_s \omega^{A(2s-2)} \, y_A...y_A\,,
\end{align}
where $y^A$ is an auxiliary commuting ``spinor''. The kinematic algebra of HS-SDYM is the polynomial current algebra of a given (color) Lie algebra $\mathfrak{g}$, i.e. $\mathbb{C}[y^A]\otimes \mathfrak{g}$. This simply means that each $\omega^{A(2s-2)}$ takes values in the adjoint representation of $\mathfrak{g}$. The field strength is the standard Yang-Mills curvature for this algebra
\begin{align}
\begin{aligned}
        F&= d\omega +[\omega,\omega]=\sum_s F_{A(2s-2)}^a T_a \,y^A...y^A=\\
     &=T_a\left(\sum_s d\omega^a_{A(2s-2)}+ \sum_{i+j=s+1} f^a_{bc}\, \omega^{b}_{A(2i-2)}\wedge \omega^c_{A(2j-2)} \right)\,y^A...y^A\,,
\end{aligned}
\end{align}
where $T_a$ are the generators of $\mathfrak{g}$. The negative helicity fields $\Psi^{A(2s)}_a$ are assumed to take values in the dual space $\mathfrak{g}^*$, which is canonically a $\mathfrak{g}$-module and there is an invariant pairing between the two. The action of HS-SDYM is \cite{Krasnov:2021nsq}
\begin{align}
    S&= \sum_s \tfrac{1}{(2s)!}\int \Psi^{A(2s)}_a H_{AA} \wedge F_{A(2s-2)}^a\,.
\end{align}
One can rewrite it in a more compact form by introducing the following pairing $\langle f|g\rangle = f_a(\pl) g^a(y)|_{y=0}$ for two generating functions, one in $\mathfrak{g}$ and another one in $\mathfrak{g}^*$.  The action reads
\begin{align}
    S&= \int\left \langle \Psi(y)| \tfrac12 H_{AA}y^A y^A \wedge F(y)\right\rangle\,.
\end{align}

\paragraph{Berends--Giele recursion.} The equations of motion for the positive helicity fields generalize \eqref{SDYMeom} and read 
\begin{align}
    \tfrac12 H_{AA}y^A y^A \wedge \left(d\omega(y) + [\omega(y),\omega(y)]\right)&=0\,.
\end{align}
It is convenient to eliminate the redundant components by setting $\Theta=0$. Then, the equations of motion for the physical fields are \cite{Guarini:2026vds}
\begin{align}\label{HSSDYMBG}
    y^A \pl_{AA'}\Phi(y)^{A'}&= \Phi(y)^{C'}\Phi(y)_{C'}\,,
\end{align}
where $\Phi(y)^{A'}=\sum_s \Phi^{A(2s-1),A'}\, y_A...y_A$ and the Lie algebra labels are made implicit. There is a simple generating function for the polarization tensors\footnote{As it is, it contains fermionic fields as well, which can be projected out by taking the $y$-odd component. } 
\begin{align}
    \Phi(y)^{A'}&=\frac{(yr)}{(rk)-(yr)} \brk^{A'}=\frac{(yr)}{(rk)}\sum_n \frac{(yr)^n}{(rk)^n} \brk^{A'}\,.
\end{align}
To make contact with SDYM, let us note that \eqref{HSSDYMBG} is very similar to \eqref{SDYMBG}. To be precise, it is easy to see that the entire dependence on $y$ is always of the form $(yr)$, i.e. all $A$-indices are carried by $r^A$. The initial polarization tensors \eqref{polarizPhi}
\begin{align}
    \Phi^{A(2s-1),A'}&= \frac{r^A...r^A}{(rk)^{2s-2}} \frac{r^A \bar{k}^{A'}}{( rk)} \,,
\end{align}
are those of SDYM but dressed with $r^A/(rk)$ to the power $2s-2$. This pattern extends to all orders. For example, the three-point functions are just those of SDYM with an HS-dressing factor (with the momentum conservation stripped off)
\begin{align}
    \mathcal{A}_3=\frac{( r3)^{2s_1-2}}{( r1)^{2s_1-2}}\frac{( r3)^{2s_2-2}}{( r2)^{2s_2-2}} \frac{( r3)^{2} (\brA \brB)}{(r1)(r2)}=\frac{( r3)^{2s-2}}{( r1)^{2s_1-2}( r2)^{2s_2-2}}\frac{( r3)^{2} (\brA \brB)}{(r1)(r2)}\,.
\end{align}
Note that in HS-SDYM the cubic interaction is such that $s_1+s_2-1=s$, where $s_{1,2}$ are the spins of two $\Phi$ and $s$ is the spin of $\Psi$. With $\delta^4(\sum_i k_i)$ added and massaged in the same way as before we get, for example, for the three-point
\begin{align}
    \mathcal{A}_3=\frac{( r3)^{2s-2}}{( r1)^{2s_1-2}( r2)^{2s_2-2}}\frac{( r3)^{4}\, \sign(\brA \brB)}{(r1)(r2)}\,\delta(13)\delta(23)\,\delta^2(\sum_k ( r k) \bar{k}_{A'})\,,
\end{align}
and four-point (s-channel explicitly) amplitudes 
\begin{align}\label{4pt_HS-SDYM}
    \mathcal{A}_4=\frac{( r4)^{2s-2}}{\prod_{i=1}^3 ( ri)^{2s_i-2}}\frac{(r4)^5}{(r1)(r2)(r3)} \sign(\brA \brB)\sign(\brC \brD) \,\delta (14)\delta (24)\delta (34)\, \delta^2(\sum_k ( r k) \bar{k}_{A'}) +t\,.
\end{align}
Therefore, it is clear that HS-SDYM has the same $n$-point amplitudes as SDYM but dressed with a factor per each positive helicity leg (associated with $\Phi$-fields):
\begin{align*}
   \text{one-leg}:\frac{( rn)^{2s_i-2}}{( ri)^{2s_i-2}}& &&\longrightarrow &&   \text{all legs}: \frac{( rn)^{2s-2}}{\prod_{i=1}^{n-1} ( ri)^{2s_i-2}}\,,
\end{align*}
where the spin of the last, negative helicity, leg is $s=\sum_i (s_i-1)+1$. In particular, taking into account the results of \cite{Guevara:2026qzd}, we conclude that HS-SDYM has the same amplitudes as SDYM dressed with the factor displayed above. This gives an example of a higher-spin theory in flat space with nonvanishing amplitudes!\footnote{Let us also mention \cite{Tran:2022amg}, where it was shown that Weinberg's low energy theorem is more about the number of derivatives in a vertex rather than the spin. In particular, gauge (like in HS-SDYM) and gravitational interactions have the usual low-derivative form in terms of chiral/twistor fields, which is not the case in terms of symmetric tensors $\Phi_{\mu_1...\mu_s}$ introduced in \cite{Fronsdal:1978rb}.} The point of the rest of the paper is to generalize this story to include all SD-theories.

\section{Self-dual (higher-spin) theories}
\label{sec:SDall}
For the sake of simplicity, but not for the sake of manifest Lorentz invariance, let us define SD-theories in the light-cone gauge. Within the Hamiltonian framework one has to build the charges of the Poincare algebra, where the Hamiltonian $H$ would determine the evolution and other generators are there to prove Lorentz invariance of the S-matrix. Most of the Poincare's algebra generators are simple and the only relation that needs to be checked at the classical level is $[H,J^{i-}]=0$. With details found in \cite{Bengtsson:1983pd,Bengtsson:1986kh,Metsaev:1991mt,Metsaev:1991nb,Ponomarev:2016lrm,Ponomarev:2022vjb}, we review the general result. Since the interactions do not contain any light-cone time derivatives, Hamiltonian $H$ and Lagrangian $\mathcal{L}$ are related in the usual way and the latter reads\footnote{It is useful to rearrange the coordinates $x^\mu$ as $x^+,x^-,x,\bar{x}$ with the metric $ds^2\sim dx^+dx^-+dxd\bar{x}$. Switching to the two-component spinor language, the coordinates are arranged into a two-by-two matrix $x^{AA'}$ with derivative $\pl_{AA'}$. A pair of such coordinates, say $x^-,x$ is denoted below $x^A$ with derivatives $\pl_A$ following the usual convention of the two-component spinor language as in Penrose-Rindler \cite{penroserindler}, e.g. $\pl^Ax^B=\epsilon^{AB}$.}
\begin{align}
    \mathcal{L}&= \sum_{s\geq0} \Phi_{-s}\square \Phi_{+s} +\sum_{\lambda_{1,2,3}} C_{\lambda_1,\lambda_2,\lambda_3}V_{\lambda_1,\lambda_2,\lambda_3} +...
\end{align}
Here, $\mathcal{L}$ is generously assumed to have fields of all spins (and even of arbitrary multiplicity, which is not indicated explicitly). Up to the cubic level this is the most general ansatz for any theory with massless fields in $4d$. 
In $4d$, massless fields have two degrees of freedom for $s>0$ and can be represented by two complex ``scalar'' fields $\Phi_{\pm s}(x)$ that are complex conjugates of each other. The kinetic term is standard and, by default, contains all spins.

The vertices are confined for now to the cubic ones. As is well-known for any triplet of helicities such that 
$\sum_i\lambda_i>0$ there is a unique vertex/three-point amplitude \cite{Bengtsson:1986kh,Benincasa:2011pg}:
\begin{align}\label{genericV}
   V_{\lambda_1,\lambda_2,\lambda_3} \sim 
    [12]^{\lambda_1+\lambda_2-\lambda_3}[23]^{\lambda_2+\lambda_3-\lambda_1}[31]^{\lambda_1+\lambda_3-\lambda_2}\,.
\end{align}
It is fixed by Lorentz invariance and there is a similar one when $\sum_i \lambda_i<0$ with the square brackets replaced by the angle ones (here we temporarily switched to the canonical spinor-helicity notation). Therefore, the entire information about the theory is encoded in the coupling constants $C_{\lambda_1,\lambda_2,\lambda_3}$ that switch on/off particular interactions and control their relative strength.

Let us assume that only vertices with positive total helicity can be turned on, i.e. $\sum_i\lambda_i>0$ in all the vertices. This can be taken as a possible definition of a self-dual theory:\footnote{Note that the  $\sum_i\lambda_i>0$ constraint eliminates $\Phi^3_0$. One should manually turn off the scalar cubic self-coupling since it is neither holomorphic nor anti-holomorphic. In fact, known higher-spin extensions force this coupling to be zero. However, there are some low spin solutions \cite{Ren:2022sws} that satisfy celestial OPE associativity while having $\Phi^3$. Let us stick here to the simplest scenario where $(\Phi_0)^3$ is absent.} a SD-theory is a Lorentz-invariant theory with $\sum_i\lambda_i>0$ interactions.\footnote{There are a few exceptions: certain chiral Lorentz-invariant theories are not self-dual; see \cite{Serrani:2025owx,Serrani:2025oaw}.}

It is in solving the closure condition $[H,J^{i-}]=0$, which is one equation for two functionals $H$ and $J^{i-}$, where locality is crucial.\footnote{Indeed, one can always solve for $J^{a-}$ provided locality is abandoned, see e.g. \cite{Ponomarev:2016lrm}. } Simple solutions are: (i) SDYM where $C_{+1,+1,-1}=g$ is the only nonvanishing coupling and color factors are implicit; (ii) SDGR where $C_{+2,+2,-2}=l_p$ is the only nonvanishing coupling ($l_p$ is the Planck length); (iii) Chiral HiSGRA where 
\cite{Metsaev:1991mt,Metsaev:1991nb,Ponomarev:2016lrm} 
\begin{align}\label{eq:magicalcoupling}
    C_{\lambda_1,\lambda_2,\lambda_3}=\frac{\kappa\,(l_p)^{\lambda_1+\lambda_2+\lambda_3-1}}{\Gamma(\lambda_1+\lambda_2+\lambda_3)}\,.
\end{align}
Note that Chiral HiSGRA has fields of all helicities and features all possible cubic interactions with $\sum_i\lambda_i>0$, i.e. it is the maximal theory of this kind.

There are two other simple solutions --- HS-extensions of SDYM and SDGR \cite{Ponomarev:2017nrr}, HS-SDYM and HS-SDGR, where $\sum_i\lambda_i=1$ and $\sum_i\lambda_i=2$, but the couplings follow the general formula above with these restrictions.\footnote{In other words, since the sum $\sum \lambda_i$ is fixed, all couplings are equal, which is closely related to Weinberg's universality, see \cite{Ponomarev:2016lrm}.} To give the simplest example, the SDYM equations of motion for $\Phi_{+1}$ read ($f^a_{bc}$ define a Lie algebra and $\pl_{A,A'}$ is a derivative in the two-component spinor language)
\begin{align}
    \square \Phi_{+1}^a + f^a_{bc}\, \partial_{A+'} \Phi_{+1}^b \partial_\fud{A}{+'} \Phi_{+1}^c&=0\,.
\end{align}
Here, some indices are explicitly $+'$ due to the light-cone gauge. One can make this choice less obvious by introducing another ``reference spinor'' $\bar{q}^{A'}$. Optimizing $D_A\equiv \bar{q}^{A'} \partial_{AA'}$, SDGR equations read
\begin{align}
    \square \Phi_{+2} + D_A D_B \Phi_{+2} D^A D^B \Phi_{+2} =0\,.
\end{align}
The vertex factor of SDYM defines a Poisson bracket, $\{f,g\}=D_Af D^Ag$. It gets ``squared'' in SDGR \cite{Monteiro:2011pc}, which is a sign of the double-copy relations \cite{Bern:2010ue}. HS-SDYM and HS-SDGR are simple HS-extensions. In HS-SDYM one keeps only vertices with $\lambda_1+\lambda_2+\lambda_3=1$ in the action
\begin{align}
    \square \Phi_{\lambda_1+\lambda_2-1}^a + f^a_{bc}\sum_{\lambda_{1,2}} D_A \Phi_{\lambda_1}^b D^A \Phi_{\lambda_2}^c&=0\,.
\end{align}
In HS-SDGR one keeps $\lambda_1+\lambda_2+\lambda_3=2$
\begin{align}
    \square \Phi_{\lambda_1+\lambda_2-2} + \sum_{\lambda_{1,2}} D_A D_B \Phi_{\lambda_1} D^A D^B \Phi_{\lambda_2} =0\,.
\end{align}
It is assumed in the two formulas above that $\lambda_1+\lambda_2$ is fixed in each equation. One variation of the same story is that one can restrict the interactions to be just $(++-)$, i.e. one can restrict $\lambda_{1,2}>0$ and, hence, $\lambda_3<0$. For these two theories there are simple covariant and gauge-invariant actions \cite{Krasnov:2021nsq},\footnote{Now, we start to see a part of the Zoo of SD-theories, see \cite{Ponomarev:2016lrm, Ponomarev:2017nrr, Krasnov:2021nsq,Monteiro:2022xwq,Serrani:2025owx}. In particular, we refer to all theories with only gauge or only gravitational interactions as HS-SDYM and HS-SDGR.} which are generalizations of those for SDYM and SDGR, respectively. HS-SDYM was already discussed in Section \ref{sec:HSSDYM}. 

For other HiSGRA applications that contain genuine higher-spin higher derivative interactions, the Poisson bracket should rather be interpreted as the first term in a star-product
\begin{align}\label{MW}
    f\star g&= \sum_k  \frac{\hbar^k}{k!}(D_{A_1}...D_{A_k}f)(D^{A_1}...D^{A_k}g)=\sum_k  \frac{\hbar^k}{k!}\{f,g\}^k
\end{align}
that deforms the point-wise (associative) multiplication along the Poisson bracket $f\star g= fg+ \hbar\{f,g\}+...$. One cannot just insert the star-product as $\Phi_\lambda \star \Phi_\lambda$ as it violates Lorentz invariance.\footnote{In fact, such models were studied, e.g. the Moyal deformation of SDYM and SDGR, see \cite{Strachan:1992em,Bu:2022iak,Bittleston:2023bzp}, and they violate Lorentz symmetry because all terms of the star-product hit the fields with a fixed helicity.} Already in the SDYM vs. SDGR examples, we see that replacing $\{\bullet,\bullet\}$ with $\{\bullet,\bullet\}^2$ forces one to change the helicity from one in SDYM to two in SDGR. In general, a necessary condition for a theory to be Lorentz invariant is to balance the number of derivatives by the total helicity: $\#D=2\sum_i\lambda_i$ in each vertex.\footnote{Let us note that the number of derivatives in a covariant formulation usually corresponds to the number of transverse derivatives in the light-cone gauge. Since $D_A\bullet D^A\bullet$ are contracted via $\epsilon^{AB}$, and hence, one of the indices is always $+'$, only one transverse derivative appears in each such contracted pair. This explains the factor of $2$. }

Taking into account the simple relation here-above between the number of derivatives (the ``power'' of the Poisson bracket) and the total helicity, the equations of motion of the most general SD-theory can now be written as
\begin{align}\label{eomgeneral}
   \delta \Phi_{\lambda_1}:& && \square \Phi_{-\lambda_1} +  \sum_{\lambda_1+\lambda_{2}+\lambda_3>0} C_{\lambda_1,\lambda_2,\lambda_3}\{\Phi_{\lambda_2},  \Phi_{\lambda_3}\}^{\sum_i \lambda_i} =0\,,
\end{align}
where the summation is over $\lambda_{2,3}$ and the equation is obtained by varying with respect to $\Phi_{\lambda_1}$, which explains $\Phi_{-\lambda_1}$ in the first term. The Lorentz invariance imposes a certain constraint on couplings $C_{\lambda_1,\lambda_2,\lambda_3}$. This constraint can also be reformulated with the help of the concept of ``kinematic algebra'' \cite{Ponomarev:2017nrr}, see \cite{Serrani:2025oaw} for further developments. The kinematic algebra is defined by extracting one contraction $\{\bullet,\bullet\}$ from the vertex, i.e. by dividing the SDYM kinematic part out.\footnote{Note that this is different from the usual kinematic algebra in the context of the double-copy. There is a number of closely related ideas in the literature starting from \cite{Monteiro:2011pc}, but we rely mostly on \cite{Ponomarev:2017nrr}, and its extension to other cases \cite{Serrani:2025oaw}.} To be precise, let us introduce auxiliary ghost variables $\Cg_\lambda(x^A)$ for fields $\Phi_\lambda(x^A,x^+,\bar{x})$. Here, $x^A\equiv x^{AA'}\bar{l}_{A'}$, $\bar{q}^{B'}\bar{l}_{B'}=1$ and, hence, $D^Ax^B=\epsilon^{AB}$. Note that only $x^A$ will matter, the rest of the coordinates are passive. The main statement is \cite{Ponomarev:2017nrr,Serrani:2025oaw}: the Lorentz invariance is equivalent to the existence of a Lie algebra  
\begin{align}
    Q^2&=0\,, && Q=\sum_{\lambda_1+\lambda_{2}+\lambda_3>0} C_{\lambda_1,\lambda_2,\lambda_3} \int d^2x \, \{\Cg_{\lambda_2},  \Cg_{\lambda_3}\}^{-1+\sum_i \lambda_i}\frac{\delta}{\delta \Cg_{-\lambda_1}}\,.
\end{align}
Here, a nilpotent ``BRST'' charge $Q$ is built from the vertices of the SD-theory with one power of $\{\bullet,\bullet\}$ chopped off, i.e. the kinematic part of the SDYM vertex is dropped. Note that due to $\lambda_1+\lambda_2+\lambda_3>0$ all vertices contain at least one ``power'' of $\{\bullet,\bullet\}$. Alternatively, one could introduce $B_\lambda$-ghost, $[\Bg^\lambda(x),\Cg_\mu(y)]_+=\delta^2(x-y)\delta_{\mu}^{\lambda}$ as a replacement for ${\delta}/{\delta \Cg_{\lambda_1}}$.

A different form of the same statement, presented first in \cite{Ponomarev:2017nrr}, is achieved with the help of a condensed notation. Firstly, the field space has a natural pairing (here, a DeWitt-type notation can be helpful with $\aA=\{\lambda\in\mathbb{Z}, \vec{p}, ...\}$ comprising the helicity, $3d$-momentum and, possibly, internal labels, if present)
\begin{align}
    (\Phi,\Psi)&=\sum_{\lambda} \int d^3p\, \Phi_{-\lambda}(-p) \Psi_{+\lambda}(+p)=\sum_{\lambda} \int d^3x\, \Phi_{-\lambda}(x) \Psi_{+\lambda}(x)\equiv \Phi^{\aA} \Psi^{\aB} G_{\aA\aB}\,.
\end{align}
The metric on the field space is ``off-diagonal'' and non-degenerate. Note that the kinetic term requires $\Phi_{+\lambda}$ and $\Phi_{-\lambda}$ to take values in the spaces dual to each other, if internal labels are present.\footnote{The scalar field is dual to itself, i.e. there is some metric $g_{\alpha\beta}$ present in the kinetic term $g_{\alpha\beta} \Phi_0^\alpha \square \Phi^\beta_0$. } Raising/lowering indices flips the sign of both the momentum and helicity, e.g. $\Phi^\aA\equiv \Phi_\lambda(p)$, $\Phi_{\aA}\equiv \Phi_{-\lambda}(-p)$. The action and the associated equations of a SD-theory can be written as
\begin{align}
    S&= \tfrac12\Phi^{\aA} \square \Phi^{\aB} G_{\aA\aB}+\tfrac1{3}\Phi^{\aA} \Phi^{\aB}\Phi^{\aC} h_{\aA\aB\aC}\,, && \square \Phi^\aA+ h\fud{\aA}{\aB\aC} \Phi^{\aB}\Phi^{\aC}=0\,.
\end{align}
Extracting the SDYM factor $\Phi_{\aA} \{\Phi^{\aB},\Phi^{\aC}\} \fA\fud{\aA}{\aB\aC}=\Phi^{\aA} \Phi^{\aB}\Phi^{\aC} h_{\aA\aB\aC}$, 
the Lorentz invariance is now equivalent to Jacobi identity for the kinematic algebra with structure constants $\fA\fud{\aA}{\aB\aC}$. It can be shown \cite{Ponomarev:2017nrr,Serrani:2025oaw} that the kinematic algebra implies that the four-point amplitude vanishes. It is also equivalent to the associativity of the celestial OPE \cite{Serrani:2025oaw}.

Therefore, behind any SD-theory, there is a Lie algebra! Let us reconsider the examples given before from the point of view of the kinematic algebra \cite{Ponomarev:2017nrr}. In SDYM the kinematic algebra is just the color Lie algebra $f^a_{bc}$ and does not have any ``kinematics'' in it. To be precise, the nilpotent charge is 
\begin{align}
    Q&= \int d^2x\,f^a_{bc}\, \Cg^b_{+1}(x)\Cg^c_{+1}(x) \Bg_a^{+1}(x)\,,
\end{align}
but $Q^2=0$ relies on $f^a_{bc}$ being the structure constants of a Lie algebra. In HS-SDYM the mechanism is the same \cite{Ponomarev:2017nrr} 
\begin{align}
    Q&= \sum_{\lambda_1+\lambda_2+\lambda_3=1}\int d^2x\,f^a_{bc}\, \Cg^b_{\lambda_2}(x)\Cg^c_{\lambda_3}(x) \Bg_a^{-\lambda_1}(x)\,.
\end{align}
If we forget about $x$, the algebra is the loop algebra $\mathbb{C}[z,z^{-1}]\otimes \mathfrak{g}$.\footnote{A convenient way to package all helicities into a single field is to define a formal Laurent series \cite{Ponomarev:2017nrr} $\Phi(p;z)=\sum_\lambda \Phi_\lambda(p)z^{\lambda-1}$. Raising/lowering DeWitt $A,B,C,...$-indices is then $z^{-2}\Phi(-p;z^{-1})$. Here, $z^{-1}$ corresponds to the scalar field and $z^0$ to spin-one.} It also works to take $\mathbb{C}[z]\otimes \mathfrak{g}$, which constrains $\lambda_{2,3}>0$. There are many other options, see \cite{Ponomarev:2017nrr,Serrani:2025owx}.\footnote{To give an infinite class that was not considered \cite{Ponomarev:2017nrr, Serrani:2025owx}, let us take a finite set $\{p_i\}$ of positive integers with $\gcd(p_1,\dots,p_k)=1$ and use them as generators $z^{p_i}$. This gives a numerical semigroup. The exponents that appear correspond to linear combinations $x=\sum_i m_i p_i$, $m_i\in \mathbb{N}$. The set of all such $x$ determines the spectrum. There is a number $M$, the Frobenius number of the semigroup, such that $x$ covers all $\mathbb{N}$ greater than $M$.} In SDGR the kinematic algebra is the Poisson algebra $\{\bullet,\bullet\}$. Again, a more precise statement is that
\begin{align}
    Q&= \int d^2x\, \{\Cg_{+2}(x),\Cg_{+2}(x)\} \Bg^{+2}(x)\,,
\end{align}
is nilpotent. In HS-SDGR, the trick is the same as in HS-SDYM  \cite{Ponomarev:2017nrr}
\begin{align}
    Q&= \sum_{\lambda_1+\lambda_2+\lambda_3=2}\int d^2x\, \{\Cg_{\lambda_2}(x),\Cg_{\lambda_3}(x)\} \Bg^{-\lambda_1}(x)\,.
\end{align}
The underlying Lie algebra is $\mathbb{C}[z,z^{-1}]\otimes \mathcal{P}$, where $\mathcal{P}$ is for the Poisson algebra. Similarly to HS-SDYM, there is a truncation $\mathbb{C}[z]\otimes \mathcal{P}$ that keeps only $(++-)$-vertices. The same infinite class of examples as in the HS-SDYM cases, which are based on numerical semigroups, can be constructed.  

In Chiral HiSGRA the underlying Lie algebra is the Moyal-Weyl commutator that originates from the associative Moyal-Weyl star-product \eqref{MW}. The latter realizes the product in the Weyl algebra, we denote it $A_1$, in terms of symbols of operators. In the gauged Chiral HiSGRA (called gluonic in \cite{Monteiro:2022xwq}) the underlying Lie algebra originates from an associative algebra that is the tensor product of the matrix algebra\footnote{Higher-spin algebras that realize Yang-Mills gaugings were studied in \cite{Konstein:1989ij} and the idea of ``gauging'' is very similar to the Chan-Paton method. In the light-cone gauge Yang-Mills gaugings of higher-spin interactions were considered in \cite{Metsaev:1991nb,Skvortsov:2020wtf}. } $\mathrm{Mat}_N$ and the star-product algebra (any associative algebra gives a Lie algebra via commutators). The nilpotent charge is
\begin{align}
    Q&= \sum_{\lambda_{1,2,3}}\int d^2x\, \frac{1}{\Gamma(\lambda_1+\lambda_2+\lambda_3)}
    \{\Cg_{\lambda_2}(x),\Cg_{\lambda_3}(x)\}^{-1+\sum_i \lambda_i} \Bg^{-\lambda_1}(x)\,.
\end{align}
In the case of matrix factors (anti)-ghosts $\Cg_\lambda$, $\Bg^\lambda$ are matrix-values and the trace is implied on the r.h.s. of $Q$. 

It turns out that there are many more SD-theories. A complete (in some sense) classification is available \cite{Serrani:2025owx} for those that have either only gauge or only gravitational interactions, i.e. $\sum_i\lambda_i=1$ and $\sum_i\lambda_i=2$, respectively. These theories are contained in HS-SDYM and HS-SDGR. All of the SD-theories are contained in Chiral HiSGRA since it has all possible $\sum_i\lambda_i>0$ vertices turned on. 

A very simple consequence of the coupling constants \eqref{eq:magicalcoupling} is that Chiral HiSGRA's amplitudes contain those of SDYM and SDGR as a subset. For example, let us take a Chiral HiSGRA with say $U(N)$-symmetry gauged. If the external helicities are restricted to be $(+...+-)$, where $\pm=\pm1$, then no higher-spin states actually propagate inside tree-level Feynman diagrams. It can be readily checked that the $(+...+-)$-amplitudes as well as BG-currents are those of SDYM.\footnote{In order to cover the SDGR case, one needs the Chiral HiSGRA with even spins only, which is formally the case of a $O(1)$-gauged theory. The spectrum of all integer spins is the case of a $U(1)$-gauged theory.}

\section{Amplitudes in SD-theories}
\label{sec:BG}
We are going to show that the kinematic algebra of a generic SD-theory allows one to reduce all tree-level amplitudes to those of SDYM times kinematic prefactors. As a trivial consequence, the same is true for the Berends--Giele currents. The BG-recursion was already applied to Chiral HiSGRA in \cite{Skvortsov:2020wtf} to show that all tree-level amplitudes vanish (for generic kinematics, now we should say in view of \cite{Guevara:2026qzd}), which did indicate that effectively the BG-currents are those of SDYM times some higher-spin (nonsingular) kinematic prefactors. In this section, the use of the light-cone gauge is mandatory, as it is the most convenient way to treat all SD theories in a unified spacetime formalism (another way would be to use twistors \cite{Tran:2021ukl,Herfray:2022prf,Tran:2022tft,Mason:2025pbz}).

In gauge theories, one usually works with the color-ordered Feynman rules by representing $f_{abc}=\Tr(T_a[T_b,T_c])$ for some generators $T_a$ of a given Lie algebra $\mathfrak{g}$. However, we cannot do that for a generic SD-theory since the kinematic algebra does not have to be the commutator Lie algebra of some associative algebra and can be a genuine Lie algebra.\footnote{There also is no guarantee that a generic kinematic algebra admits a non-degenerate invariant quadratic form. For example, the Poisson algebra on polynomial functions does not. } There are cases, e.g. the full Chiral HiSGRA, where the Lie bracket is a commutator in an associative algebra. 

A natural alternative to the standard color ordering is the DDM ordering \cite{DelDuca:1999iql,DelDuca:1999rs}, which, in fact, has many advantages compared to the usual color ordering via $T_a$, e.g. the Kleiss-Kuijf relations \cite{Kleiss:1988ne} are resolved. The DDM basis, which can be defined for any Lie algebra, is given by ``ladders'' of the form 
\begin{align}\notag
    L\fdu{a_1 a_{\sigma_2}...a_{\sigma_{n-1}}}{a_n}&=f\fdu{a_1 a_{\sigma_2}}{x_1}f\fdu{x_1 a_{\sigma_3}}{x_2} ... f\fdu{x_{n-3}  a_{\sigma_{n-1}}}{a_n}=
    \parbox{5cm}{\ladder}
\end{align}
The ladders correspond to nested Lie brackets of the simplest form $[...[[e_{a_1},e_{a_{\sigma_2}}],e_{a_{\sigma_3}}], ..., e_{a_{\sigma_{n-1}}}]^{a_n}$. 
The distinguishing feature is the simple topology of the contracted structure constants. It was shown in \cite{DelDuca:1999rs} that all possible contractions of $f^{a}_{bc}$ can be reduced to this basis with the help of Jacobi identities, anti-symmetry of $f^a_{bc}$ and nothing else.\footnote{The proof in \cite{DelDuca:1999rs} did appeal to the usual color ordering, but is expected to be independent of this relation. } Also, the Feynman rules are the same as for the usual color ordering: one has to draw all planar graphs, which correspond to ``colorless'' Feynman diagrams. The complete amplitude can be represented as 
\begin{align}
    \mathcal{A}_{12...n}&=\sum_{S_{n-2}} L\fdu{a_1 a_{\sigma_2}...a_{\sigma_{n-1}}}{a_n} \tilde{\mathcal{A}}_n(1, \sigma_2,...,\sigma_{n-1},n)\,.
\end{align}
Here, $\tilde{\mathcal{A}}_n$ is a partial amplitude built from Feynman graphs (stripped of the color factors) that are planar in the given ordering of the external legs. Since any SD-theory is SDYM $\times$ (some) kinematic algebra and the structure constants of the kinematic algebra obey the Jacobi identity and, in general, nothing else, the DDM-ordering can be directly applied to any SD-theory.\footnote{Perhaps one of the first instances in the literature where it was explicitly stated, i.e. that one can use kinematic algebras in the DDM basis, is \cite{Johansson:2015oia}, but see also \cite{Bern:2011ia,Fu:2018hpu} for closely related ideas and \cite{Bjerrum-Bohr:2012kaa} for the study of SDGR as SDYM squared. }  This way, we get the desired result both for amplitudes and, if needed, for the BG-recursion as the latter is just a recursive way to compute the same amplitude. In our case, partial amplitudes $\tilde{\mathcal{A}}_n$ are those of SDYM. 

Since it sounds too simple to be true, let us give a few examples to convince the reader. For SDYM we just get the DDM-ordering for the color:
\begin{align}
    \mathcal{A}_{12...n}^{\text{SDYM}}&=\sum_{S_{n-2}} L\fdu{a_1 a_{\sigma_2}...a_{\sigma_{n-1}}}{a_n} \tilde{\mathcal{A}}^{\text{SDYM}}_n(1, \sigma_2,...,\sigma_{n-1},n)\,.
\end{align}
The partial amplitudes of SDYM are gauge invariant. Outside the scope of \cite{Guevara:2026qzd}, i.e. for real Minkowski kinematics, SDYM amplitudes vanish. The gauge invariance of the amplitudes of \cite{Guevara:2026qzd} follows from the collinearity of the physical momenta and from the fact that the amplitudes have zero weight with respect to the reference spinor $r^A$. Since the DDM ladders depend on kinematics for a generic (SD-)theory, one cannot conclude that each term in the sum is gauge invariant, except for HS-SDYM.

\paragraph{Example: HS-SDYM.} This is almost the case of SDYM: the kinematic algebra does not have anything kinematic in it except for helicity labels. This case was already considered in Section \ref{sec:HSSDYM} and let us reframe it in the DDM-setting.  It is convenient to pass to $\hat\lambda=\lambda-1$ as it removes $1$ from $\lambda_1+\lambda_2-\lambda_3=1$, where the sign of $\lambda_3$ is flipped in accordance with the rule to get $\fA^{A}_{BC}$. Indeed, we find that $C^{\hat\lambda_1}_{\hat\lambda_2,\hat\lambda_3}=\delta^{\hat\lambda_1}_{\hat\lambda_2+\hat\lambda_3}$. At the four-point level the DDM-factor is 
\begin{align}   
f\fdu{a_1 a_{2}}{x_1}f\fdu{x_{1}  a_{{3}}}{a_4}&\times \left(\sum_{\hat\lambda}C^{\hat\lambda}_{\hat\lambda_1,\hat\lambda_2}C^{\hat\lambda_4}_{\hat\lambda, \hat\lambda_3}=\delta^{\hat\lambda_4}_{\hat\lambda_1+\hat\lambda_2+\hat\lambda_3}\right)\,,
\end{align}
i.e. it is the color DDM-factor times a constraint on the helicities. Note that, given the external helicities, the helicity of all inner lines is fixed in HS-SDYM. At the $n$-point level we simply have the product of the DDM-color factor 
\begin{align}
    L\fdu{a_1 a_{\sigma_2}...a_{\sigma_{n-1}}}{a_n} \times \delta^{\hat\lambda_n}_{\hat\lambda_1+\hat\lambda_{\sigma_2}+...+\hat\lambda_{\sigma_{n-1}}}\,,
\end{align}
and the Kronecker symbols that determine the helicity of the last leg. In the version of HS-SDYM restricted to $(++-)$-vertices, one needs to additionally require $\lambda_{1,...,n-1}>0$, i.e. $\hat{\lambda}_{1,...,n-1}\geq0$. In what follows, we will use a more standard and compact notation for the DDM-ladders $\mathbf{c}(1,\sigma,n)$ instead of $L\fdu{a_1 a_{\sigma_2}...a_{\sigma_{n-1}}}{a_n}$.

\paragraph{Light-cone gauge interlude.} To compute amplitudes in the light-cone gauge we have to introduce some notation for the momentum-space counterparts of the Poisson bracket. As in position space, everything nontrivial is confined to the $2d$ subspace. A generic $4$-momentum $p^\mu$ is $(p,\bar{p},p^+\equiv \beta, p^-)$. The Lorentz-invariance confines the transverse components $p$ and $\bar{p}$ to appear only in the form of 
\begin{align}
    &\PPb_{ij}=\bar{p}_i\beta_j-\bar{p}_j\beta_i\,,&
&\PP_{ij}=p_i\beta_j-p_j\beta_i\,,
\end{align}
where $i,j,...$ refers to external legs. We can choose $\PPb$ to be exactly the Fourier image of $\{\bullet,\bullet\}$ in our coordinates. It is a product of two spinors, e.g. $\bar{i}=(\bar{p}_i,-\beta_i)$, etc., and $\PPb_{ij}=(\bar{i}\bar{j})$. At the three-point level, thanks to momentum conservation, one finds $\PPb=\PPb_{12}= \PPb_{23}=\PPb_{31}$. The kinematic part of the SDYM vertex in momentum space is simply $\PPb$. It transforms in the anti-symmetric representation of the permutation group $S_3$. More light-cone notation can be found in Appendix \ref{app:LC}. The most general vertex that leads to equations \eqref{eomgeneral} has the following form in momentum space
\begin{align}
    \Phi^{\aA} \Phi^{\aB}\Phi^{\aC} h_{\aA\aB\aC}=\sum_{\lambda_{1,2,3}} C_{\lambda_1,\lambda_2,\lambda_3}\int \delta^3(\sum_i p_i) \,f_{a_1a_2a_3}\Phi_{\lambda_1,p_1}^{a_1}\Phi_{\lambda_2,p_2}^{a_2}\Phi_{\lambda_3,p_3}^{a_3}\,\PPb^{\lambda_1+\lambda_2+\lambda_3}\,.
\end{align}
Here, Latin indices $a_{1,2,3}$ span the internal space, if present, at a given helicity. The internal space can depend on helicity, e.g. the $\lambda=+1$ field can take values in the adjoint of some Lie algebra $\mathfrak{g}$, while fields of other helicities can take values in various representations of $\mathfrak{g}$. Strictly speaking, $f_{a_{\lambda_1},a_{\lambda_2},a_{\lambda_3}}$ would be a more precise notation. We will avoid it as we will deal only with three cases: (i) all fields take values in the adjoint of $\mathfrak{g}$, which is equipped with a non-degenerate bilinear invariant form, in which case we can simply write $\Tr(\Phi_{\lambda_1,p_1}\Phi_{\lambda_2,p_2}\Phi_{\lambda_3,p_3})$; (ii) all fields $\Phi_\lambda$ take values in $\mathrm{Mat}_N$, where notation $\Phi^a_\lambda T_a$ can be appropriate; (iii) trivial internal space. It is convenient to think that $C_{\lambda_1,\lambda_2,\lambda_3}$ is defined for all $\lambda_{1,2,3}$ and, in particular, it vanishes whenever $\sum \lambda_i\leq0$. With these precautions, the structure constants of the kinematic algebra are read off from
\begin{align}\notag
\Phi_{\aA} \{\Phi^{\aB},\Phi^{\aC}\} \fA\fud{\aA}{\aB\aC}&= \sum_{\lambda_{1,2,3}} C_{-\lambda_1,\lambda_2,\lambda_3}\int \delta^4(\sum_i p_i) \,f_{a_1a_2a_3}\Phi_{-\lambda_1,-p_1}^{a_1}\Phi_{\lambda_2,p_2}^{a_2}\Phi_{\lambda_3,p_3}^{a_3}\,\PPb^{-\lambda_1+\lambda_2+\lambda_3-1}\, \PPb\,,   
\end{align}
where the momentum conservation has $(-p_1+p_2+p_3)$, in fact, and the single $\PPb$ on the right is to factor out one Poisson bracket.

Without taking \cite{Guevara:2026qzd} into account, i.e. with the naive propagators and/or Minkowski kinematics, the four-point amplitude can be shown to be proportional to the Jacobi identity of the kinematic algebra \cite{Serrani:2025oaw} 
\begin{align}
\nonumber
\mathcal{A}_4=\,&\mathcal{A}_s+\mathcal{A}_t+\mathcal{A}_u=
    \sum_{\omega}\mathcal{F}_{1234\omega}\PPb_{12}^{\lambda_{12}+\omega}\frac{1}{(p_1+p_2)^2}\PPb_{34}^{\lambda_{34}-\omega}+2\leftrightarrow 4+2\leftrightarrow 3\\
    =\,&\frac{\PPb_{12}\PPb_{34}}{(p_1+p_2)^2}\sum_{\omega}\Big(\mathcal{F}_{1234\omega}\PPb_{12}^{\lambda_{12}+\omega-1}\PPb_{34}^{\lambda_{34}-\omega-1}+\mathcal{F}_{2314\omega}\PPb_{23}^{\lambda_{23}+\omega-1}\PPb_{14}^{\lambda_{14}-\omega-1}\\
    \nonumber
    &+\mathcal{F}_{3124\omega}\PPb_{31}^{\lambda_{13}+\omega-1}\PPb_{24}^{\lambda_{24}-\omega-1}\Big)\sim \fA\fdu{A_1A_2}{B}\fA\fdu{BA_3}{A_4}+\fA\fdu{A_2A_3}{B}\fA\fdu{BA_1}{A_4}+\fA\fdu{A_3A_1}{B}\fA\fdu{BA_2}{A_4}=0\,,
\end{align}
where $\mathcal{F}_{1234\omega}\equiv C^{\lambda_1,\lambda_2,\omega}C^{-\omega,\lambda_3,\lambda_4}f\fdu{a_1a_2}{c}f\fdu{ca_3}{a_4}$ and $\lambda_{ij}\equiv\lambda_i+\lambda_j$. In order to factor out the term $\sim\PPb\PPb$, we used the on-shell relations in Appendix \ref{app:identities}. The vanishing of the naive four-point amplitude for all self-dual theories is a consequence of the Jacobi identity for the kinematic algebra \cite{Ponomarev:2017nrr,Serrani:2025oaw}.

A slightly more general result is a three-point BG-current, i.e. the four-point amplitude with one leg kept off-shell. Assuming the Jacobi identity for the kinematic algebra, we find 
\begin{align}
\nonumber
\mathcal{A}_4&=\frac{\PPb_{12}\PPb_{34}}{s_{12}}\fA\fdu{A_1A_2}{B}\fA\fdu{BA_3}{A_4}+\frac{\PPb_{23}\PPb_{14}}{s_{23}}\fA\fdu{A_2A_3}{B}\fA\fdu{BA_1}{A_4}+\frac{\PPb_{31}\PPb_{24}}{s_{31}}\fA\fdu{A_3A_1}{B}\fA\fdu{BA_2}{A_4}\\
\nonumber
&=\left(\frac{\PPb_{12}\PPb_{34}}{s_{12}}-\frac{\PPb_{23}\PPb_{14}}{s_{23}}\right)\fA\fdu{A_1A_2}{B}\fA\fdu{BA_3}{A_4}+\left(\frac{\PPb_{31}\PPb_{24}}{s_{31}}-\frac{\PPb_{23}\PPb_{14}}{s_{23}}\right)\fA\fdu{A_3A_1}{B}\fA\fdu{BA_2}{A_4}\\
\nonumber
&=\prod\limits_{i=2}^4\beta_i\left(\frac{\beta_3}{4\PP_{34}\PP_{23}}\fA\fdu{A_1A_2}{B}\fA\fdu{BA_3}{A_4}+\frac{\beta_2}{4\PP_{24}\PP_{23}}\fA\fdu{A_3A_1}{B}\fA\fdu{BA_2}{A_4}\right)p_1^2\\
&=\left(\frac{1}{\PP^{\beta}_{34}\PP^{\beta}_{23}}\fA\fdu{A_1A_2}{B}\fA\fdu{BA_3}{A_4}+\frac{1}{\PP^{\beta}_{24}\PP^{\beta}_{23}}\fA\fdu{A_3A_1}{B}\fA\fdu{BA_2}{A_4}\right)\,p_1^2\,,
\end{align}
where we used the off-shell relations \eqref{offshell1} and \eqref{offschell2}. The last line features two DDM-factors at the four-point level. 

The general expression for a BG-current in SDYM is well-known \cite{Berends:1987me,Bardeen:1995gk,Krasnov:2016emc}. The $n$-point partial amplitude for SDYM with one leg off-shell reads 
\begin{equation}\label{SDYM-npoint}
    \tilde{\mathcal{A}}^{\text{SDYM}}_n(1^-2^+\cdots n^+)=\frac{(-)^n\beta_3\cdots \beta_{n-1}\,p_1^2}{2^{n-2}\PP_{23}\PP_{34}\cdots \PP_{n-1,n}}\prod\limits_{i=2}^n\beta_i=\frac{(-)^np_1^2}{\PP^{\beta}_{23}\PP^{\beta}_{34}\cdots \PP^{\beta}_{n-1,n}}\,.
\end{equation}
This is the usual color-ordered partial amplitude/BG-current, which appears both in the standard color-ordered decomposition and in the DDM-ordered one.

If a given SD-theory can be thought of as coming from an associative algebra, i.e. the kinematic algebra is the commutator Lie algebra, the standard color-ordered decomposition can be more handy. Let us assume that $T_a$ are generators of a color algebra, i.e. we have $\Phi_\lambda^aT_a$. The $n$-point amplitude then reads
\begin{align}\label{trace_decomposition}
    \begin{split}
    \mathcal{A}_n^{\text{tree}}&=\sum_{\sigma\in S_{n-1}}C_n(1,\sigma_2,\cdots,\sigma_n)\,\tilde{\mathcal{A}}_n^{\text{SDYM}}(1,\sigma_2,\cdots, \sigma_n)\,\Tr(T_{a_1}T_{a_{\sigma_2}}\cdots T_{a_{\sigma_n}})\,,
    \end{split}
\end{align}
where we defined
\begin{equation}\label{A_1...n_LC}
C_n(1,2,...,n)=\sum_{\omega_i}\mathcal{C}^{\lambda_1,\lambda_2,\cdots,\lambda_n}\PPb_{12}^{\lambda_{12}+\omega_1-1}\PPb_{<3}^{\lambda_3+\omega_2-\omega_1-1}\cdots\PPb_{n-1,n}^{\lambda_{n-1}+\lambda_n-\omega_{n-3}-1}\,.
\end{equation}
Due to the nature of SD-theories, the sum over $\omega_i$ runs over all integer values such that the $\PPb_{ij}$ always appear with nonnegative powers. We have also introduced the following shorthand notation\footnote{To make the contact with the underlying associative structure explicit, we can define the generators to be $t_a(p)=E_pT_a$, where $E_p=e^{ipx}$ is a plane-wave; the star-product is defined by $E_{p_1}\star E_{p_2}=E_{p_1+p_2}\exp[\hbar \PP_{12}]$. Then, $t_a(p)$ represent the generators of the tensor product of two associative algebras. One can naturally introduce the bookkeeping $z$, via $t_{a,\lambda}(p)=z^{\lambda-1} E_pT_a$ and $\hbar=l_p/z$. The trace $\Tr$ consists of (1) the trace on the color part $T_a$; (2) integral over $x$, which gives the overall momentum conservation; (3) picking the coefficient of $z^{-2}$. With these definitions we find, $\Tr(t_{a,\lambda}(p_1)\star t_{b,\mu}(p_2))=\delta_{\mu+\lambda,0}\delta^4(p_1+p_2)\Tr(T_a T_b)$. More generally, the associative product reads $ 
t_{a,\lambda}(p_1)\star t_{b,\mu}(p_2)
=
\sum_{k\ge0}
\frac{\bigl(l_p\PPb_{12} \bigr)^k}{k!}
z^{\lambda+\mu-2-k}\,T_aT_bE_{p_1+p_2}$.}
\begin{align}\label{C-notation}
   &\mathcal{C}^{\lambda_1,\lambda_2,\cdots,\lambda_n}\equiv C^{\lambda_1,\lambda_2,\omega_1}C^{-\omega_1,\lambda_3,\omega_2}\cdots C^{-\omega_{n-3},\lambda_{n-1},\lambda_n}\,,&
    &\PPb_{<j}=\sum_{i<j}\PPb_{ij}\,.
\end{align}

Otherwise, if the kinematic algebra is a genuine Lie algebra, rather than the commutator Lie algebra derived from an associative one, we use the DDM-decomposition \cite{DelDuca:1999rs}:
\begin{align}\label{DDM_decomposition}
    \begin{split}
    \mathcal{A}_n^{\text{tree}}&=\sum_{\sigma\in S_{n-2}}\left(\fA\fdu{A_1A_{\sigma_2}}{B_1}\fA\fdu{B_1A_{\sigma_3}}{B_2}\cdots \fA\fdu{B_{n-3}A_{\sigma_{n-1}}}{A_n}\right)\tilde{\mathcal{A}}_n^{\text{SDYM}}(1,\sigma_2,\cdots, \sigma_{n-1},n)\\
    &=\sum_{\sigma\in S_{n-2}}\mathbf{c}(1,\sigma,n)\tilde{\mathcal{A}}_n^{\text{SDYM}}(1,\sigma,n)\,,
    \end{split}
\end{align}
where the sum runs over $(n-2)!$ permutations of the color factors $\mathbf{c}(1,\sigma,n)$ and we defined $\tilde{\mathcal{A}}_n^{\text{SDYM}}(1,\sigma,n)\equiv \tilde{\mathcal{A}}_n^{\text{SDYM}}(1,\sigma_2,\cdots, \sigma_{n-1},n)$. The relation between this basis and the trace decomposition is made explicit by the Kleiss–Kuijf relations. In the decompositions above, $\tilde{\mathcal{A}}_n^{\text{SDYM}}$ corresponds to the $n$-point partial amplitude for SDYM, i.e. the sum over all color-ordered planar diagrams; this is a gauge invariant quantity. On the other hand, the kinematic algebra factor $\mathbf{c}(1,\sigma,n)$ may not be gauge invariant. The total amplitude, after summing over all permutations, will then be a gauge invariant object.

\paragraph{Example: Chiral HiSGRA. } Let us reproduce the result of \cite{Skvortsov:2020wtf}, where all tree-level amplitudes and BG-currents of Chiral HiSGRA have been computed. In this case, we should use the trace decomposition \eqref{trace_decomposition}. A specific example is $U(N)$ Chiral HiSGRA, where all cubic vertices, both even- and odd-derivative, are activated with couplings fixed to the special values in \eqref{eq:magicalcoupling}. 
In this case, we have 
\begin{equation} C_n(1,2,...,n)=(l_p)^{\Lambda_n-(n-2)}\sum_{\omega_i}\frac{\PPb_{12}^{\lambda_{12}+\omega_1-1}\PPb_{<3}^{\lambda_3+\omega_2-\omega_1-1}\cdots\PPb_{<n-2}^{\lambda_{n-2}+\omega_{n-3}-\omega_{n-4}-1}\PPb_{n-1,n}^{\lambda_{n-1}+\lambda_n-\omega_{n-3}-1}}{\Gamma(\lambda_{12}+\omega_1)\cdots\Gamma(\lambda_{n-1}+\lambda_n-\omega_{n-3})}\,.
\end{equation}
By an appropriate change of variables dictated by 
\begin{equation}
    \lambda_{12}+\omega_1-1=k_1\,,\quad
    \lambda_3+\omega_2-\omega_1-1=k_2\,,\quad
    \cdots\quad
    \lambda_{n-1}+\lambda_n-\omega_{n-3}-1=k_{n-2}\,,
\end{equation}
and noting that we have
\begin{equation}
    \sum_{i=1}^{n-2}k_i=\sum_{i=1}^n\lambda_i-(n-2)\,.
\end{equation}
Using the multinomial formula, we find 
\begin{equation}
    C_n(1,2,...,n)=\frac{(l_p)^{\Lambda_n-(n-2)}\alpha_n^{\Lambda_n-(n-2)}}{(\Lambda_n-(n-2))!}\,,\qquad
    \Lambda_n=\sum_{i=1}^{n}\lambda_i\,,\quad
    \alpha_n\equiv \sum_{i<j}^{n-2}\PPb_{ij}+\PPb_{n-1,n}\,.
\end{equation}
The total amplitude then becomes
\begin{align}\label{chiral_HISGRA_npt_LC}
    \begin{split}
    \mathcal{A}_n^{\text{tree}}&=\sum_{\sigma\in S_{n-1}}\frac{(l_p)^{\Lambda_n-(n-2)}\alpha_{n,\sigma}^{\Lambda_n-(n-2)}}{(\Lambda_n-(n-2))!}\tilde{\mathcal{A}}_n^{\text{SDYM}}(1,\sigma_2,\cdots, \sigma_n)\mathrm{Tr}(T_{a_1}T_{a_{\sigma_2}}\cdots T_{a_{\sigma_n}})\,,
    \end{split}
\end{align}
where $\alpha_{n,\sigma}$ is the appropriate color-ordered version of $\alpha_{n}$. This coincides with the result presented in \cite{Skvortsov:2020wtf} upon plugging in the expression \eqref{SDYM-npoint} for the SDYM amplitudes with one off-shell leg.

For chiral higher-spin theory without the $U(N)$ gauging, one allowed theory consists of having all couplings fixed to the value \eqref{eq:magicalcoupling}, but only for even derivative vertices. All odd derivative vertices must be set to zero. Therefore, following a similar procedure as for the $U(N)$ case, we get
\begin{equation}
    C_n(1,2,...,n)=(l_p)^{\Lambda_n-(n-2)}\sum_{k_i\,\text{odd}}\frac{\PPb_{12}^{k_1}\PPb_{<3}^{k_2}\cdots\PPb_{n-1,n}^{k_{n-2}}}{k_1!k_2!\cdots k_{n-2}!}\,,\qquad \sum_{i=1}^{n-2}k_i=\Lambda_n-(n-2)\,.
\end{equation}
This can be summed into 
\begin{align}
C_n
&=\frac{(l_p)^{\Lambda_n-(n-2)}}
{2^{n-2}(\Lambda_n-(n-2))!}
\sum_{\epsilon_2=\pm1}\cdots\sum_{\epsilon_{n-1}=\pm1}
\left(\prod_{j=2}^{n-1}\epsilon_j\right)
\left[
\sum_{j=2}^{n-2}\epsilon_j\sum_{i<j}\PPb_{ij}
+\epsilon_{n-1}\PPb_{n-1,n}
\right]^{\Lambda_n-(n-2)}
\nonumber\\
&=\frac{(l_p)^{\Lambda_n-(n-2)}}
{2^{n-3}(\Lambda_n-(n-2))!}
\sum_{\epsilon_2=\pm1}\cdots\sum_{\epsilon_{n-2}=\pm1}
\left(\prod_{j=2}^{n-2}\epsilon_j\right)
\left[
\sum_{j=2}^{n-2}\epsilon_j\sum_{i<j}\PPb_{ij}
+\PPb_{n-1,n}
\right]^{\Lambda_n-(n-2)}\,,
\end{align}
where the sum over the various $\epsilon_i$ projects the sum into odd powers only. Equivalently, introducing an auxiliary variable $t$, the odd-power projection can be written compactly as
\begin{equation}
C_n
=(l_p)^{\Lambda_n-(n-2)}
[t^{\,\Lambda_n-(n-2)}]
\left[
\prod_{j=2}^{n-2}
\sinh\left(t\sum_{i<j}\PPb_{ij}\right)
\right]
\sinh\left(t\PPb_{n-1,n}\right)\,,
\end{equation}
where $[t^K]F(t)$ denotes the coefficient of $t^K$ in the expansion of
$F(t)$. This can be shown using 
\begin{equation}
\sinh(tX_j)
=
\frac{1}{2}
\sum_{\epsilon_j=\pm1}
\epsilon_j\,e^{t\epsilon_j X_j}\,,\qquad
[t^K]\,e^{tY}
=
\frac{Y^K}{K!}\,.
\end{equation}

\paragraph{Example: (HS)-SDGR.} SDGR is a very instructive example, see e.g. \cite{Krasnov:2016emc,Hasuwannakit:2025agr}. Since SDGR does not have any associative algebra behind it, we have to resort to the DDM-decomposition. For example, for SDGR at the four-point level, there are the usual $s$, $t$, and $u$ diagrams. The DDM color factors have the form $f\fdu{a_1 a_{\sigma_2}}{x_1}f\fdu{x_1  a_{\sigma_3}}{a_4}$, i.e. there are only two. The $u$-diagram can be reduced via the Jacobi identity to have the same color factors as the first two, see below. For higher-point amplitudes, the DDM-ladder contains only simple kinematics and is 
\begin{align}\label{An_SDGR}
    \mathbf{c}(1,2,...,n)&=(l_p)^{n-2}\PPb_{12}\PPb_{<3}\cdots \PPb_{<n-2}\PPb_{n-1,n}=(l_p)^{n-2}\prod_{j=2}^{n-2}\left(\sum_{i<j}\PPb_{ij}\right)\PPb_{n-1,n}\,.
\end{align}
For HS-SDGR and numerous other theories \cite{Ponomarev:2017nrr,Krasnov:2021nsq,Serrani:2025owx} we find simply
\begin{align}\label{An_HS-SDGR}
\mathbf{c}(1,2,...,n)=\mathcal{C}^{\lambda_1,\lambda_2,\cdots,\lambda_n}\prod_{j=2}^{n-2}\left(\sum_{i<j}\PPb_{ij}\right)\PPb_{n-1,n}\,.
\end{align}

\subsection{Half-collinear amplitudes}
The main point of the discussion here-above is that amplitudes and BG-currents of SD-theories can be reduced to those of SDYM times products of the structure constants of a given kinematic algebra. It is important that the kinematic factors are non-singular. Single-minus gluon tree amplitudes have been shown to be nonzero in the half-collinear limit \cite{Guevara:2026qzd}, see also \cite{Witten:2003nn,Roiban:2004yf}.

Below, we give some examples of non-vanishing amplitudes in the light-cone gauge and match the results of \cite{Guevara:2026qzd,Guevara:2026qwa}. We consider mostly SDYM, as for all other self-dual theories the amplitudes can be recovered via either the usual color decomposition or the DDM decomposition. However, an example of SDGR can be instructive for the reader.

In the following, all external momenta are on shell, so we write $p_{iAA'}=i_A\bar{i}_{A'}$. In the normalization of Appendix~\ref{app:LC}, the undotted spinor bracket is $(ij)=2\PP_{ij}/(\beta_i\beta_j)\equiv\PP^\beta_{ij}$. In particular, $\PP^\beta_{ir}\equiv(ir)=-\epsilon^{AB}i_A r_B$ for any reference spinor $r_A$, which need not correspond to an external leg. Below, $r_A$ and $\mu_A$ are chosen with $(r\mu)\neq0$, so that the two projections together impose full momentum conservation, as in Appendix~\ref{app:identities}.

\paragraph{SDYM.} In SDYM the cubic color-stripped amplitude is
\begin{align}
\begin{split}
\mathcal{A}_3&=\PPb_{12}\delta^4\left(\sum_i i_A\bar{i}_{A'}\right)=\PPb_{12}\delta^2\left(\sum_i\PP^{\beta}_{ri}\bar{i}_{A'}\right)\delta^2\left(\sum_i\PP^{\beta}_{i\mu }\bar{i}_{A'}\right)(\PP^{\beta}_{r \mu})^2\\
&=\PPb_{12}\delta^2\left(\sum_i\PP^{\beta}_{ri}\bar{i}_{A'}\right)\delta^2\left(\PP^{\beta}_{13}\bar{1}_{A'}+\PP^{\beta}_{23}\bar{2}_{A'}\right)(\PP^{\beta}_{r3})^2\\
&=\text{sg}_{12}\,\delta(\PP^{\beta}_{13})\delta(\PP^{\beta}_{23})\delta^2\left(\sum_i\PP^{\beta}_{ri}\bar{i}_{A'}\right)(\PP^{\beta}_{r3})^2\,.
\end{split}
\end{align}
From now on we use the notation:
\begin{equation}
\delta^{02}\equiv\delta^2\left(\sum_i\PP^{\beta}_{ri}\bar{i}_{A'}\right)\,.
\end{equation}
Here, $\text{sg}_{12}=\sign(\PPb_{12})$ and, more generally, $\text{sg}_{i,jk}=\sign(\PPb_{ij}+\PPb_{ik})$, etc. For the four-point function, we consider the s-channel cut where the internal propagator is replaced by delta-functions. We start with the $s$-channel:
\begin{align}\notag
\mathcal{A}_s&=\PPb_{12}\delta(p_{34}^2)\PPb_{34}\delta^4\left(\sum_i i_A\bar{i}_{A'}\right)=\text{sg}_{34}\delta(\PP^{\beta}_{34})\PPb_{12}\delta^2(\PP^{\beta}_{14}\bar{1}_{A'}+\PP^{\beta}_{24}\bar{2}_{A'}+\cancel{\PP^{\beta}_{34}\bar{3}_{A'}})\delta^{02}(\PP^{\beta}_{r4})^2\\
    &=\text{sg}_{34}\text{sg}_{12}\,\delta(\PP^{\beta}_{14})\delta(\PP^{\beta}_{24})\delta(\PP^{\beta}_{34})\delta^{02}(\PP^{\beta}_{r4})^2\,.
\end{align}
For the $t$-channel instead we get 
\begin{align}  \mathcal{A}_t&=\PPb_{23}\delta(p_{41}^2)\PPb_{41}\delta^4\left(\sum_i i_A\bar{i}_{A'}\right)=\text{sg}_{41}\text{sg}_{23}\,\delta(\PP^{\beta}_{14})\delta(\PP^{\beta}_{24})\delta(\PP^{\beta}_{34})\delta^{02}(\PP^{\beta}_{r4})^2\,.
\end{align}
The total four-point amplitude is
\begin{equation}
    \mathcal{A}_4=\mathcal{A}_s+\mathcal{A}_t=(\text{sg}_{34}\text{sg}_{12}+\text{sg}_{41}\text{sg}_{23})\delta(\PP^{\beta}_{14})\delta(\PP^{\beta}_{24})\delta(\PP^{\beta}_{34})\delta^{02}(\PP^{\beta}_{r4})^2\,,
\end{equation}
which is the same as \cite{Guevara:2026qzd} through the light-cone gauge lenses.

The $5$-point amplitude is more complicated. Let us consider a representative example that corresponds to the maximal cut of one of the five planar cubic diagrams. For the maximal cut, the propagators on internal lines are replaced by $\delta(p^2)$:
\begin{equation}
\begin{aligned}
\begin{tikzpicture}[baseline=-0.5ex]

\def\R{1}        
\def\n{5}        
\def\dn{3}       
\def\leg{0.5}    

\draw (0,0) circle (\R);

\foreach \k in {1,...,\n} {
  \coordinate (p\k) at ({-360/\n*(\k-1+\dn)}:\R);
  \node at ({-360/\n*(\k-1)}:{\R+\leg+0.05}) {$p_{\k}$};
}

\coordinate (a) at ({-360/\n*(1.5-1+\dn)}:{\R*0.4});
\coordinate (b) at ({-360/\n*(1.5-1+\dn)}:{\R*0});
\coordinate (c) at ({-360/\n*(4.5-1+\dn)}:{\R*0.4});

\draw (p1) -- (a) -- (p2) ;
\draw (p3) -- (b);
\draw (c) -- (p4) ;
\draw (c) -- (p5) ;

\draw[red, very thick] (a) -- (b) -- (c);

\end{tikzpicture}
\qquad \mathcal{A}^{\text{max  cut}}_5=\PPb_{23}\delta(p_{23}^2)\PPb_{45}\delta(p_{45}^2)(\PPb_{12}+\PPb_{13})\delta^{02}(\PP^{\beta}_{r\mu})^2\delta^2(\sum_i\PP^{\beta}_{i\mu}\bar{i}_{A'})\,.
\end{aligned}
\end{equation}
Then step by step we get
\begin{align}
    \begin{split}
    \mathcal{A}^{\text{max  cut}}_5&=\text{sg}_{23}\text{sg}_{45}\,\delta(\PP^{\beta}_{23})\delta(\PP^{\beta}_{45})(\PPb_{12}+\PPb_{13})\delta^{02}(\PP^{\beta}_{r5})^2\delta^2(\PP^{\beta}_{15}\bar{1}_{A'}+\PP^{\beta}_{25}\bar{2}_{A'}+\PP^{\beta}_{35}\bar{3}_{A'})\\
    &=\text{sg}_{23}\text{sg}_{45}\,\delta(\PP^{\beta}_{23})\delta(\PP^{\beta}_{45})(\PPb_{12}+\PPb_{13})\delta^{02}(\PP^{\beta}_{r5})^2\delta^2\left(\PP^{\beta}_{15}\bar{1}_{A'}+\PP^{\beta}_{35}\left(\bar{2}_{A'}+\bar{3}_{A'}\right)\right)\\
    &=\text{sg}_{23}\text{sg}_{45}\,\delta(\PP^{\beta}_{23})\delta(\PP^{\beta}_{45})(\PPb_{12}+\PPb_{13})\delta^{02}(\PP^{\beta}_{r5})^2\delta(\PP^{\beta}_{15})\delta(\PP^{\beta}_{35})\frac{1}{\Big|\PPb_{12}+\PPb_{13}\Big|}\\
    &=\text{sg}_{23}\text{sg}_{45}\text{sg}_{1,23}\,\delta(\PP^{\beta}_{15})\delta(\PP^{\beta}_{25})\delta(\PP^{\beta}_{35})\delta(\PP^{\beta}_{45})\delta^{02}(\PP^{\beta}_{r5})^2\,,
    \end{split}
\end{align}
where in the second line we have used that, in the normalization \eqref{Pbeta_norm}, $\PP^\beta_{ij}=2(p_i/\beta_i-p_j/\beta_j)$, so that the identity $\PP^\beta_{25}=\PP^\beta_{23}+\PP^\beta_{35}$ holds and
\begin{equation}
    \PP^{\beta}_{23}=0\quad\implies\quad \PP^{\beta}_{25}=\PP^{\beta}_{35}\,.
\end{equation}

\paragraph{SDGR.} Using the previous results and with the help of the DDM-decomposition \eqref{DDM_decomposition} and the formula in \eqref{An_SDGR}, we can compute the $4$-point amplitude for SDGR:
\begin{align}
    \begin{split}
    \mathcal{A}_4^{\text{SDGR}}&=\mathbf{c}(1,2,3,4)\tilde{\mathcal{A}}^{\text{SDYM}}_{1234}+\mathbf{c}(1,3,2,4)\tilde{\mathcal{A}}^{\text{SDYM}}_{1324}\\
    &=(l_p)^2\left(\PPb_{12}\PPb_{34}(\text{sg}_{12}\text{sg}_{34}+\text{sg}_{23}\text{sg}_{41})+\PPb_{13}\PPb_{24}(\text{sg}_{13}\text{sg}_{24}+\text{sg}_{32}\text{sg}_{41})\right)\times\\
    &\qquad\delta(\PP^{\beta}_{14})\delta(\PP^{\beta}_{24})\delta(\PP^{\beta}_{34})\delta^{02}(\PP^{\beta}_{r4})^2\,.
    \end{split}
\end{align}
After some manipulation and use of
\begin{equation}
    \PPb_{12}\PPb_{34}=-\PPb_{31}\PPb_{24}-\PPb_{23}\PPb_{14}\,,
\end{equation}
we find 
\begin{equation}
    \mathcal{A}_4^{\text{SDGR}}=(l_p)^2\left(|\PPb_{12}||\PPb_{34}|+|\PPb_{13}||\PPb_{24}|+|\PPb_{14}||\PPb_{23}|\right)\times\delta(\PP^{\beta}_{14})\delta(\PP^{\beta}_{24})\delta(\PP^{\beta}_{34})\delta^{02}(\PP^{\beta}_{r4})^2\,.
\end{equation}
This is the correct result, as it can be easily compared to \cite{Guevara:2026qwa}. 

\section{Self-dual amplitudes in spinor-helicity language}
\label{sec:spinorhell}

In this section, we recast the nonvanishing half-collinear tree-level amplitudes of self-dual theories in spinor-helicity variables. We then study the scattering of higher-spin multiplets by summing over the external helicities and constructing the associated generating functions. The main ``leap of faith'' is to use the CSW formalism to extend on-shell cubic spinor-helicity amplitudes of (higher-spin) fields off-shell in order to compute higher-point amplitudes.    

\subsection{Self-dual cubic amplitudes}

We start from the holomorphic (self-dual) cubic amplitudes:
\begin{equation}\label{standard_cubic}
    \mathcal{A}(1_{\lambda_1}2_{\lambda_2}3_{\lambda_3})=C^{\lambda_1,\lambda_2,\lambda_3}[12]^{\lambda_{12}-\lambda_3}[23]^{\lambda_{23}-\lambda_1}[31]^{\lambda_{13}-\lambda_2}\,,\qquad \lambda_{123}>0\,,
\end{equation} 
where we define $\lambda_{ij}\equiv\lambda_i+\lambda_j$ and $\lambda_{ijk}\equiv\lambda_i+\lambda_j+\lambda_k$. Using momentum conservation, we can bring it to the following form:
\begin{align}\label{general_cubic}
\begin{split}
    \mathcal{A}(1_{\lambda_1}2_{\lambda_2}3_{\lambda_3})&=C^{\lambda_1,\lambda_2,\lambda_3}[12]^{\lambda_{12}-\lambda_3}[23]^{\lambda_{23}-\lambda_1}[31]^{\lambda_{13}-\lambda_2}\\
    &=C^{\lambda_1,\lambda_2,\lambda_3}\frac{(r1)^x(r2)^y}{(r3)^{x+y}}[12]^{\lambda_{12}-\lambda_3+x+y}[23]^{\lambda_{23}-\lambda_1-x}[31]^{\lambda_{13}-\lambda_2-y}\,,
\end{split}
\end{align}
where $x,y\in \mathbb{Z}$ and we used the identities
\begin{align}\label{momentum_conservation}
    &1_A\bar{1}_{A'}+2_A\bar{2}_{A'}+3_A\bar{3}_{A'}=0&
    &\implies&
    &[jk]=\frac{(ri)}{(rj)}[ki]\,,
\end{align}
consequences of momentum conservation, and where $i,j,k$ are distinct and cyclically ordered.

\paragraph{From light-cone to spinor-helicity.} The same result can be found starting from the light-cone holomorphic cubic vertices, proceeding as follows. The holomorphic cubic vertices in the light-cone gauge are
\begin{equation}\label{holo_cubic_LC}
    H^{\lambda_1,\lambda_2,\lambda_3}_3=C^{\lambda_1,\lambda_2,\lambda_3}\int d^9q \,\delta\Big(\sum_i q_i\Big)h_3\Phi^{\lambda_1}_{q_1}\Phi^{\lambda_2}_{q_2}\Phi^{\lambda_3}_{q_3}\,,\quad
    h_3=\frac{\PPb_{12}^{\lambda_{123}}}{\beta_1^{\lambda_1}\beta_2^{\lambda_2}\beta_3^{\lambda_3}}\,,\qquad \lambda_{123}>0\,.
\end{equation}
On-shell $(p_i^-=-p_i\bar p_i/\beta_i)$, the relation between the light-cone and spinor-helicity is clear, up to the little-group scaling:
\begin{align}
    &\frac{1}{\sqrt{2}}\,p_{iAA'}=
    \begin{pmatrix}
        -\frac{p_i\bar{p}_i}{\beta_i} & p_i\\
        \bar{p}_i & - \beta_i \end{pmatrix}=i_A\bar{i}_{A'}\,,&
    &i_A=\sqrt{\omega_i}(z_i,1)\,,&
    &\bar{i}_{A'}=-\sqrt{\omega_i}(\bar{z}_i,1)\,,
\end{align}
where
\begin{align}
&\omega_i=\beta_i\,,&
&z_i=-\frac{p_i}{\beta_i}\,,&
&\bar{z}_i=-\frac{\bar{p}_i}{\beta_i}&
&\implies&
&i_A=\frac{1}{\sqrt{\beta_i}}(-p_i,\beta_i)\,,&
&\bar{i}_{A'}=\frac{1}{\sqrt{\beta_i}}(\bar{p}_i,-\beta_i)\,.
\end{align}
The brackets are defined as in Appendix \ref{app:LC}, then 
\begin{align}
&(ij)=-\epsilon^{AB}i_Aj_B\,,&
&[ij]=-\epsilon^{A'B'}\,\bar{i}_{A'}\bar{j}_{B'}\,.
\end{align}
Therefore, we find 
\begin{align}
&[ij]=\frac{\PPb_{ij}}{\sqrt{\beta_i\beta_j}}\,,&
&(ij)=\frac{\PP_{ij}}{\sqrt{\beta_i\beta_j}}\,,&
&\PPb_{ij}=\bar{p}_i\beta_j-\bar{p}_j\beta_i\,,&
&\PP_{ij}=p_i\beta_j-p_j\beta_i\,.
\end{align}
Remember that even if we always retain the freedom to perform little-group scaling, the particular choice of spinors we have adopted to represent the momenta is the symmetric one. This choice directly affects how the light-cone expressions map onto the standard spinor-helicity form of the cubic amplitudes. One can verify that this is a choice for which the light-cone cubic vertices reproduce exactly the standard spinor-helicity cubic amplitudes. 

By choosing the reference spinor $r=(1,0)$, then $(ri)=-\sqrt{\beta_i}$, we find the relations:
\begin{equation}\label{fromLCtoSH}
\PPb_{ij}=(ri)(rj)[ij]\,,\qquad
\PP_{ij}=(ri)(rj)(ij)\,,\qquad
\beta_i=(ri)^2\,.
\end{equation}
Starting from \eqref{holo_cubic_LC}, we get 
\begin{equation}\label{standard2_cubic}
    h_3=\frac{\PPb_{12}^{\lambda_{123}}}{\beta_1^{\lambda_1}\beta_2^{\lambda_2}\beta_3^{\lambda_3}}\quad
    \longrightarrow\quad\mathcal{A}_3=
    \frac{[12]^{\lambda_{123}}}{(r1)^{\lambda_1-\lambda_{23}}(r2)^{\lambda_2-\lambda_{13}}(r3)^{2\lambda_3}}
\end{equation}
These amplitudes match \eqref{general_cubic} for $x=\lambda_{23}-\lambda_1$, and $y=\lambda_{13}-\lambda_2$.

The identification established above was achieved for a particular choice $r=(1,0)$. However, one can verify that the amplitude is, in fact, independent of the choice of the reference spinor $r$. Consequently, the resulting expression for the amplitude is valid for an arbitrary choice of the reference spinor. In particular, the following relations between the spinor-helicity and light-cone gauge formalisms can be identified:
\begin{itemize}

    \item 
    A particular choice of light-cone gauge corresponds to a choice of a null reference spinor in the spinor-helicity formalism (see also \cite{Boels:2013bi,Monteiro:2022xwq}).

    \item 
    The off-shell rescaling of $\Phi_\lambda(p)\to t^{2\lambda}\Phi_\lambda(p)$ preserves the free action and changes the vertices off-shell. On shell, its only effect is to change the normalisation of the external states, which is exactly little-group scaling.
    
\end{itemize}
The essential distinction between the two formalisms is that the light-cone formulation is defined off-shell, whereas the spinor-helicity formalism is intrinsically on-shell. In this sense, the light-cone variables provide an off-shell continuation of the spinor-helicity kinematics, as discussed in \cite{Ponomarev:2016cwi}. On the other hand, the spinor-helicity formalism has the advantage of making Lorentz covariance manifest.

The first statement follows immediately from \eqref{fromLCtoSH}. The second statement follows from the following observation. An on-shell Lorentz invariant amplitude must transform under little-group rescalings as ($\bar{i}_{A'}\rightarrow t_i\bar{i}_{A'}\,,\,i_A\rightarrow t_i^{-1}i_A$) according to
\begin{equation} \mathcal{A}_n((t^{-1}_11_A,t_1\bar{1}_{A'})_{\lambda_1},...,(t^{-1}_nn_A,t_n\bar{n}_{A'})_{\lambda_n})=\prod_{i=1}^n t^{2\lambda_i}_i\mathcal{A}_n((1_A,\bar{1}_{A'})_{\lambda_1},...,(n_A,\bar{n}_{A'})_{\lambda_n})\,. 
\end{equation}
The transformed amplitude is expressed in terms of a different choice of spinor representatives for the same external momenta. Although little-group scaling leaves the momenta invariant, it rescales the external wavefunctions and hence the normalization of the corresponding one-particle states. In the light-cone formalism, the same effect is reproduced by an appropriate rescaling of the fields. Under
\begin{align}\label{LC_fieldred} 
&\Phi_{q_i}^{\lambda_i}\rightarrow t^{2\lambda_i}_i\Phi_{q_i}^{\lambda_i}& 
&\implies& &h_n(t^{2\lambda_1}_1\Phi_{q_1}^{\lambda_1},...,t^{2\lambda_n}_n\Phi_{q_n}^{\lambda_n})=\prod_{i=1}^n t^{2\lambda_i}_ih_n(\Phi_{q_1}^{\lambda_1},...,\Phi_{q_n}^{\lambda_n})\,. \end{align}
Let us illustrate this explicitly for the three-point amplitude by choosing $t_i=(ri)$. After a little-group scaling, the amplitude $\mathcal{A}_3$ in \eqref{standard2_cubic} becomes
\begin{equation}\label{example_LGscaling} 
\mathcal{A}_3=(r1)^{\lambda_{123}}(r2)^{\lambda_{123}} [12]^{\lambda_{123}}\,.
\end{equation}
Starting from $h_3$ as in \eqref{standard2_cubic} and redefining the fields as in \eqref{LC_fieldred}, we get
\begin{equation}
h_3=\PPb_{12}^{\lambda_{123}}\,,
\end{equation}
and using \eqref{fromLCtoSH}, one recovers \eqref{example_LGscaling}.

\paragraph{Setting the stage for higher-point amplitudes.}

Before proceeding with the computations, let us set some notation. In the following section we will always separate the contribution of the SDYM $n$-point amplitude and the factor coming from the kinematical algebra. The $n$-point SDYM partial amplitude in spinor-helicity is
\begin{equation}\label{A_12...n}
        \tilde{\mathcal{A}}^{\text{SDYM}}_n(1^+,...,(n-1)^+,n^-)=\frac{(rn)^{n+1}}{\prod_{i=1}^{n-1}(ri)}A_{1\cdots n}\prod_{a=1}^{n-1}\delta(an)\delta^{0|2}\,,\quad
    \delta^{0|2}=\delta^2\Big(\sum_i (ri) \bri_{A'}\Big)\,,
\end{equation}
where $A_{1\cdots n}$ is the stripped SDYM chamber function that can be found in \cite{Guevara:2026qzd}. 

On the half-collinear support, the unprimed spinors are all
proportional. To make gauge invariance manifest, we combine the
factor $(rn)^2$ with the delta function $\delta^{0|2}$ to obtain
\begin{equation}
    (rn)^2\delta^2\Big(\sum_i(ri)\bar{i}_{A'}\Big)
    =\delta^2\Big(\sum_i\frac{(ri)}{(rn)}\bar{i}_{A'}\Big)
    \equiv \hat{\delta}^{0|2},
    \qquad (rn)\neq 0.
\end{equation}
We can then rewrite \eqref{A_12...n} as
\begin{equation}
    \tilde{\mathcal{A}}^{\text{SDYM}}_n=\frac{(rn)^{n-1}}{\prod_{i=1}^{n-1}(ri)}\hat{A}_{1\cdots n}\,,\qquad \hat{A}_{1\cdots n}\equiv A_{1\cdots n}\prod_{a=1}^{n-1}\delta(an)\hat{\delta}^{0|2}\,.
\end{equation}
From now on, to check gauge invariance, we only need to verify that the total homogeneity degree in $r$ is zero.

\subsection{Four-point amplitudes}
To compute higher-point amplitudes in the spinor-helicity formalism, we use the Cachazo--Svrček--Witten (CSW) off-shell prescription \cite{Cachazo:2004kj} (see also \cite{Gorsky:2005sf,Mansfield:2005yd}) to continue the holomorphic cubic amplitudes off-shell. A null momentum factorises as $K_{AA'}=k_A\bar{k}_{A'}$. For $K^2\neq 0$, we can instead write
\begin{equation}\label{notK^2=0}
K_{AA'}=k_A\bar{k}_{A'}+r_A\bar{\xi}_{A'}\,,
\end{equation}
where the second term completes the rank-one piece $k_A\bar k_{A'}$ to the off-shell momentum.

To establish this decomposition, choose a reference spinor \(r_A\) and an auxiliary spinor \(k_A\) with \((rk)\neq0\), and define
\begin{equation}\label{bark_def}
\bar{k}_{A'}\equiv\frac{r^A K_{AA'}}{(rk)}\,.
\end{equation}
It follows that
\begin{equation}
r^A\left(K_{AA'}-k_A\bar{k}_{A'}\right)=0\,.
\end{equation}
In a two-dimensional undotted spinor space, the expression in parentheses must therefore be proportional to \(r_A\), giving \eqref{notK^2=0}. In particular,
\begin{equation}
(rk)\bar{k}_{A'}=r^A K_{AA'}\,,
\end{equation}
holds even when \(K^2\neq0\). This is the only relation needed below.

The decomposition also has a light-cone interpretation \cite{Ponomarev:2016cwi}. The on-shell correspondence between the two formalisms is given in \eqref{diag_momenta}. Although the rank-one factorization of the full momentum no longer holds off-shell, the light-cone parametrisation remains well-defined and we can write
\begin{equation}\label{LC-CSW}
\frac{1}{\sqrt{2}}\,p_{AA'}=P_{AA'}+\left(p^-+\frac{p\,\bar{p}}{\beta}\right)
\begin{pmatrix}
1 & 0\\
0 & 0
\end{pmatrix}\,,
\qquad
P_{AA'}=
\begin{pmatrix}
-\frac{p\bar{p}}{\beta} & p\\
\bar{p} & -\beta
\end{pmatrix}\,.
\end{equation}
Here \(P_{AA'}\) is the on-shell parametrisation, while the remaining term carries the off-shell information, as in \eqref{notK^2=0}.

We start by computing the four-point $s$-channel amplitude. We use the CSW prescription to extend the exchanged field off-shell. Taking all external momenta outgoing and using momentum conservation
\begin{align}\label{4pt-mom}
    &1_A\bar{1}_{A'}+2_A\bar{2}_{A'}+K_{AA'}=0\,,&
    &3_A\bar{3}_{A'}+4_A\bar{4}_{A'}-K_{AA'}=0\,.
\end{align}
where $K_{AA'}$ denotes the momentum flowing out of the left vertex, we find 
\begin{equation}
(rk) \bar{k}_{A'}=-(r1)\bar{1}_{A'}-(r2)\bar{2}_{A'}\,,\qquad
(rk) \bar{k}_{A'}=(r3)\bar{3}_{A'}+(r4)\bar{4}_{A'}\,,
\end{equation}
where we contracted \eqref{4pt-mom} with a reference spinor $r_A$. These are projected momentum-conservation equations, and they do not require $K^2=0$. Now contracting the expressions above with $\bar{i}_{A'}$, for $i=1,2,3,4$, we find 
\begin{subequations}
\begin{align}\label{mom1}
(rk)[2k]&=(r1)[12]\,,
&
(rk)[k1]&=(r2)[12]\,,\\\label{mom2}
(rk)[4k]&=-(r3)[34]\,,
&
(rk)[k3]&=-(r4)[34]\,.
\end{align}
\end{subequations}
Starting from the left vertex in the form \eqref{general_cubic}
\begin{equation}
V_L=\frac{(r1)^x (r2)^y}{(rk)^{x+y}}
[12]^{\lambda_{12}-\omega+x+y}
[2k]^{\lambda_2-\lambda_1+\omega-x}
[k1]^{\lambda_1-\lambda_2+\omega-y}\,,
\end{equation}
and using \eqref{mom1} we find
\begin{equation}\label{left-cubic-vertex}
V_L=
(rk)^{-2\omega}
[12]^{\lambda_{12}+\omega}
(r1)^{\lambda_2-\lambda_1+\omega}
(r2)^{\lambda_1-\lambda_2+\omega}\,.
\end{equation}
Instead, using \eqref{mom2} in the right vertex we get
\begin{equation}
V_R=
(rk)^{2\omega}
[34]^{\lambda_{34}-\omega}
(r3)^{\lambda_4-\lambda_3-\omega}
(r4)^{\lambda_3-\lambda_4-\omega}\,.
\end{equation}
Therefore the $s$-channel exchange diagram is
\begin{align}
\begin{split}
\mathcal{A}_s
&=\sum_\omega
\mathcal F_{1234\omega}\,
V_L\,\frac1s\,V_R\\
&=\sum_{\omega} \mathcal{F}_{1234\omega}\frac{[12]^{\lambda_{12}+\omega}[34]^{\lambda_{34}-\omega}}{[12](12)}(r1)^{\lambda_2-\lambda_1+\omega}(r2)^{\lambda_1-\lambda_2+\omega}(r3)^{\lambda_4-\lambda_3-\omega}(r4)^{\lambda_3-\lambda_4-\omega}\,,
\end{split}
\end{align}
where we used the most general form for the cubic vertex with ``structure constants''  $\mathcal{F}_{1234\omega}\equiv C^{\lambda_1,\lambda_2,\omega}C^{-\omega,\lambda_3,\lambda_4}f\fdu{a_1a_2}{c}f\fdu{ca_3}{a_4}$. We can also factor out the SDYM part of the amplitude
\begin{equation}
    \mathcal{A}_s^{\text{SDYM}}=\frac{(r4)^3}{(r1)(r2)(r3)} \text{sg}[12]\text{sg}[34] \,\delta (14)\delta (24)\delta (34)\, \hat{\delta}^{0|2}\,,
\end{equation}
and we find 
\begin{align}\label{4pts-channel}
    \begin{split}
    \mathcal{A}_s&=\sum_{\omega}\mathcal{F}_{1234\omega}[12]^{\lambda_{12}+\omega-1}[34]^{\lambda_{34}-\omega-1}(r1)^{\lambda_2-\lambda_1+\omega+1}(r2)^{\lambda_1-\lambda_2+\omega+1}\\
    &\qquad\quad\qquad\quad(r3)^{\lambda_4-\lambda_3-\omega+1}(r4)^{\lambda_3-\lambda_4-\omega-3}\times\mathcal{A}^{\text{SDYM}}_s\,.
    \end{split}
\end{align}
As expected, the dependence on the auxiliary parameters $x$ and $y$ has completely cancelled. We can verify \eqref{4pts-channel} against the known result for HS-SDYM theory in the case in which only $(++-)$ vertices are present. For the general HS-SDYM theory, we have the following constraint on the cubic vertices
\begin{align}
    &\lambda_1+\lambda_2+\omega=1\,,&
    &\lambda_3+\lambda_4-\omega=1\,.
\end{align}
From \eqref{4pts-channel} we get
\begin{equation}
    \mathcal{A}_s=\frac{(r4)^{-2\lambda_4-2}}{(r1)^{2\lambda_1-2}(r2)^{2\lambda_2-2}(r3)^{2\lambda_3-2}}\times \mathcal{A}^{\text{SDYM}}_s\,.
\end{equation}
When only $(++-)$ vertices are present, this can be rewritten as
\begin{equation}
    \mathcal{A}_s=\frac{(r4)^{2s-2}}{(r1)^{2\lambda_1-2}(r2)^{2\lambda_2-2}(r3)^{2\lambda_3-2}}\times \mathcal{A}^{\text{SDYM}}_s\,,\qquad
    s=\sum_{i=1}^3(s_i-1)+1\,,
\end{equation}
and coincides with \eqref{4pt_HS-SDYM}. To recover the full amplitude, we also have to add the $t$- and $u$-channels.

As a remark, one could instead compute tree-level amplitudes directly from the covariant equations of chiral higher-spin gravity \cite{Skvortsov:2022syz,Sharapov:2022faa}, by computing Berends--Giele currents and amputating the off-shell leg, as done in \cite{Guarini:2026vds} for HS-SDYM.\footnote{Reference \cite{Guarini:2026vds} computes the all-multiplicity Berends--Giele currents of HS-SDYM restricted to the $(++-)$ sector.} It seems that the CSW continuation is more economical. However, it is not clear to which extent this off-shell extension of the on-shell vertices/amplitudes can be justified for the most general (higher-spin) case. 

The use of the CSW vertices requires more than their agreement with the light-cone vertices on shell. Off-shell vertices are not unique: two continuations with the same three-point amplitude may differ by terms proportional to the inverse propagator
\begin{equation}
    V_3=V_3^{\mathrm{CSW}}+s\,R\,.
    \label{eq:offshell-ambiguity}
\end{equation}
Sewing together two such vertices gives
\begin{equation}
    V_L\,\frac{1}{s}\,V_R=V_L^{\mathrm{CSW}}\frac{1}{s}V_R^{\mathrm{CSW}}+R_LV_R^{\mathrm{CSW}}+V_L^{\mathrm{CSW}}R_R+s\,R_LR_R\,,
\end{equation}
where the subscripts $L$ and $R$ refer to the left and right vertex. The last three terms have no pole at $s=0$: they are contact terms, and they would be missed if one simply sewed CSW vertices.

However, such terms are an artefact of the choice of field variables. A field redefinition that shifts the cubic vertex by $s\,R$ also generates quartic vertices, whose contributions cancel these contact terms, so that the amplitudes do not change.\footnote{The same mechanism is used in the light-cone gauge: there, terms proportional to $\PP\PPb$, i.e.\ to $s$, in the cubic vertices are removed by field redefinitions \cite{Metsaev:1991mt,Metsaev:1991nb}. This is what makes the separation of the cubic vertices into holomorphic ones, depending only on $\PPb$, and anti-holomorphic ones, depending only on $\PP$, manifest in the light-cone formulation.} The CSW sewing is therefore correct provided the CSW vertices are the cubic vertices of a field basis in which the theory has no quartic vertices. The light-cone formulation is such a basis, and we now show that the CSW vertices coincide with the light-cone ones off-shell.

This is easily verified on the vertices derived above. Take $r=(1,0)$, which fixes the light-cone gauge, so that \eqref{fromLCtoSH} applies: $(ri)^2=\beta_i$ and $[12]=\PPb_{12}/\big((r1)(r2)\big)$. Substituting this into \eqref{left-cubic-vertex} the powers of $(r1)$ and $(r2)$ combine into
\begin{equation}
V_L=(rk)^{-2\omega}\,\PPb_{12}^{\lambda_{12}+\omega}\,(r1)^{-2\lambda_1}(r2)^{-2\lambda_2}
=\frac{\PPb_{12}^{\lambda_{12}+\omega}}{\beta_1^{\lambda_1}\beta_2^{\lambda_2}\,(rk)^{2\omega}}\,.
\end{equation}
The auxiliary spinor $k_A$ in \eqref{bark_def} is arbitrary, and we may choose it such that $(rk)^2=\beta_K$; in any case the factors $(rk)^{\mp2\omega}$ cancel between $V_L$ and $V_R$. This gives
\begin{align}
&V_L=\frac{\PPb_{12}^{\lambda_{12}+\omega}}{\beta_1^{\lambda_1}\beta_2^{\lambda_2}\beta_K^{\,\omega}}\,,&
&V_R=\frac{\PPb_{34}^{\lambda_{34}-\omega}}{\beta_3^{\lambda_3}\beta_4^{\lambda_4}\beta_K^{-\omega}}\,,
\end{align}
which are precisely the light-cone vertices $h_3$ of \eqref{holo_cubic_LC} with helicities $(\lambda_1,\lambda_2,\omega)$ and $(-\omega,\lambda_3,\lambda_4)$, where $\PPb_{12}=\bar K\beta_1-\bar p_1\beta_K$ by momentum conservation. Since only the projected relations \eqref{mom1}--\eqref{mom2}, which are valid for $K^2\neq0$, were used, the CSW vertices coincide with the light-cone vertices off-shell: no correction $s\,R$ is present, and, since the light-cone theory has cubic vertices only, sewing CSW vertices reproduces its amplitudes exactly. As we will see the same argument applies to every vertex at higher multiplicity. The arguments above might sound plausible, but the main justification comes mostly from the final result.

\paragraph{From light-cone to spinor-helicity.} The result obtained above precisely reproduces the corresponding expression in the light-cone gauge
\begin{equation}\label{s-channel_LC}
\mathcal{A}_s=\sum_{\omega}\mathcal{F}_{1234\omega}\PPb_{12}^{\lambda_{12}+\omega-1}\PPb_{34}^{\lambda_{34}-\omega-1}\times \mathcal{A}^{\text{SDYM}}_s\,.
\end{equation}
To establish the correspondence, we proceed as follows. The amplitude in \eqref{s-channel_LC} is obtained after the field redefinition $\Phi_{q_i}^{\lambda_i}\rightarrow\beta^{\lambda_i}_i\Phi_{q_i}^{\lambda_i}$ which is the little-group scaling with $t_i=(ri)$, since $t_i^{2\lambda_i}=(ri)^{2\lambda_i}=\beta_i^{\lambda_i}$. We therefore have to compare it to \eqref{4pts-channel} after performing the little-group scaling with $t_i=(ri)$. This gives
\begin{align}
    \begin{split}
    \mathcal{A}_s=&\sum_{\omega}\mathcal{F}_{1234\omega}[12]^{\lambda_{12}+\omega-1}[34]^{\lambda_{34}-\omega-1}(r1)^{\lambda_{12}+\omega}(r2)^{\lambda_{12}+\omega}\times\\
    &(r3)^{\lambda_{34}-\omega}(r4)^{\lambda_{34}-\omega}\text{sg}[12]\text{sg}[34] \,\delta (14)\delta (24)\delta (34)\,\delta^2\Big(\sum_i\frac{(ri)}{(r4)}\bar{i}_{A'}\Big)\,,
    \end{split}
\end{align}
which, upon using \eqref{fromLCtoSH}, is in exact agreement with \eqref{s-channel_LC}.

\subsection{Higher-point amplitudes}
We can now apply the same reasoning to derive the $n$-point tree amplitude of a generic SD-theory. As discussed above, when an SD-theory originates from an associative algebra, i.e. the kinematic algebra is the commutator Lie algebra, the standard color-ordered decomposition is particularly convenient. The amplitude can then be written as
\begin{align}\label{trace/commutator}
    \begin{split}
    \mathcal{A}_n^{\text{tree}}&=\sum_{\sigma\in S_{n-1}}C_n(\sigma_1,\cdots,\sigma_{n-1},n)\,\tilde{\mathcal{A}}_n^{\text{SDYM}}(\sigma_1,\cdots,\sigma_{n-1},n)\,\Tr(T_{a_{\sigma_1}}\cdots T_{a_{\sigma_{n-1}}} T_{a_n})\,.
    \end{split}
\end{align}
When the kinematic algebra is instead a genuine Lie algebra and no preferred associative-algebra trace representation is available, it is more natural to use the DDM decomposition
\begin{equation}
\mathcal{A}_n^{\text{tree}}=\sum_{\sigma\in S_{n-2}}\mathbf{c}(1,\sigma,n)\,\tilde{\mathcal{A}}_n^{\text{SDYM}}(1,\sigma,n)\,.
\end{equation}
The derivation is straightforward once we recall that any SD-theory can be viewed schematically as a ``product’’ of a specific kinematical algebra with SDYM theory. The cubic amplitude takes the schematic form
\begin{align}
    \begin{split}
    \mathcal{A}_3
    &=C^{\lambda_1,\lambda_2,\lambda_3}\frac{(r1)^{x+1}(r2)^{y+1}}{(r3)^{x+y+2}}[12]^{\lambda_{12}-\lambda_3-1+x+y}[23]^{\lambda_{23}-\lambda_1-x}[31]^{\lambda_{13}-\lambda_2-y}\times \frac{(r3)^2}{(r1)(r2)}[12]\\
    &=\fA\fdu{A_1A_2}{A_3}\times \frac{(r3)^2}{(r1)(r2)}[12]=\fA\fdu{A_1A_2}{A_3}\times \mathcal{K}_{\text{SDYM}}\,.
    \end{split}
\end{align}
This makes explicit the separation between the kinematical part of the SDYM vertex ($\mathcal{K}_{\text{SDYM}}$) and the remaining theory-dependent factor. As emphasised previously, the SDYM factor obeys the usual Berends–Giele recursion, while the theory-dependent factors combine through contractions of the kinematical structure constants $\fA\fdu{A_1A_2}{A_3}$ along the cubic tree. 

Using either the color-ordered decomposition or the DDM decomposition, the full amplitude is therefore completely determined by the SDYM partial amplitudes $\tilde{\mathcal{A}}_n^{\text{SDYM}}$ together with the kinematical factor obtained by contracting the kinematical structure constants $\fA\fdu{A_1A_2}{A_3}$. In order to compute $C_n$ and, equivalently, $\mathbf{c}(1,\sigma,n)$ we can proceed as before. Indeed, the same CSW continuation extends the four-point computations to arbitrary multiplicity.

For each momentum $p_{AA'}$, on-shell or off-shell, we define $\chi_{A'}(p)=r^A p_{AA'}$ and introduce $[\chi(p_i)\chi(p_j)]_r=\chi(p_i)^{A'}\chi(p_j)_{A'}=(ri)(rj)[ij]$.\footnote{With our normalization, $[\chi(p_i)\chi(p_j)]_r=\PPb_{ij}$ and $(ri)^2=\beta_i$, so the light-cone vertex $h_3$ and $V_3$ coincide.} Here, $i_A$ and $\bar{i}_{A'}$ must be interpreted as spinors for on-shell momenta in the case of external legs and as in \eqref{bark_def} for internal legs. At each cubic vertex momentum conservation gives $\chi(p_i)+\chi(p_j)+\chi(p_k)=0$, and hence $[\chi(p_i)\chi(p_j)]_r=[\chi(p_j)\chi(p_k)]_r=[\chi(p_k)\chi(p_i)]_r$, irrespective of whether the momenta are on-shell. Consequently, a vertex with helicity labels $\lambda_i,\lambda_j,\lambda_k$ takes the form
\begin{equation}
    V_3=\frac{[\chi(p_i)\chi(p_j)]_r^{\lambda_{ijk}}}{(ri)^{2\lambda_i}(rj)^{2\lambda_j}(rk)^{2\lambda_k}}=\frac{((ri)(rj)[ij])^{\lambda_{ijk}}}{(ri)^{2\lambda_i}(rj)^{2\lambda_j}(rk)^{2\lambda_k}}\,.
\end{equation}
The factors $(r\ell)^{\bullet}$ on each internal edge cancel between its two cubic vertices because the helicity labels on the two ends are opposite.

Relative to the corresponding SDYM tree with external helicities $(+,...,+,-)$, this leaves the external factor
\begin{equation}
    \frac{(rn)^{-2\lambda_n-2}}
{\prod_{i=1}^{n-1}(ri)^{2\lambda_i-2}}\,,
\end{equation}
and one factor $[\chi(p_i)\chi(p_j)]_r^{\lambda_{ijk}-1}$ per vertex. Therefore, at the end, we arrive at
\begin{align}
\begin{split}\label{A_n_cov}
\mathbf{c}(1,\ldots,n)={}&C_n(1,...,n)=
\frac{(rn)^{-2\lambda_n-2}}
{\prod_{i=1}^{n-1}(ri)^{2\lambda_i-2}}
\sum_{\omega_i}
\mathcal C^{\lambda_1,\ldots,\lambda_n}
\\
&\times
[12]_r^{\lambda_{12}+\omega_1-1}
\prod_{j=3}^{n-2}
[<j]_r^{\lambda_j+\omega_{j-1}-\omega_{j-2}-1}
[n-1\,n]_r^{\lambda_{n-1}+\lambda_n-\omega_{n-3}-1}\,,
\end{split}
\end{align}
where $\mathcal{C}^{\lambda_1,\lambda_2,\cdots,\lambda_n}$ was introduced in \eqref{C-notation} and because of the structure of SD theories, the sums over the integers $\omega_i$ are restricted to values for which all brackets $[\bullet\bullet]_r$ appear with non-negative powers. We have introduced the following shorthand notation
\begin{equation}
    [<j]_r:=\sum_{i<j}[ij]_r\,,\qquad
    [ij]_r:=(ri)[ij](rj)\,.
\end{equation}
This covariant sign function was already suggested in \cite{Guevara:2026qzd}. Every weighted bracket contributes two powers of $r_A$ and the sum of the exponents of the various brackets $[\bullet\bullet]_r$ is $\sum_i\lambda_i-(n-2)$, so these powers cancel those of the external prefactor. Therefore $\mathbf{c}$, or equivalently $C_n$, has total homogeneity degree zero in $r_A$; it is therefore independent of the common reference spinor on the half-collinear support, and then gauge invariant.

As an important consistency check, one can verify that this result coincides with the light-cone expression in \eqref{A_1...n_LC}, once the two formulations are compared following the same procedure used above for the three- and four-point amplitudes.



\paragraph{Example: Chiral HiSGRA. } A specific example is the $U(N)$ Chiral HiSGRA, where all cubic vertices, both even- and odd-derivative, are activated with the same couplings $\sim \Gamma[\lambda_{123}]^{-1}$. In this case, we have 
\begin{align} 
\begin{split}
&C_n=(l_p)^{\Lambda_n-(n-2)}\frac{(rn)^{-2\lambda_n-2}}{\prod_{i=1}^{n-1}(ri)^{2\lambda_i-2}}\times\\
&\sum_{\omega_i}\frac{[12]_r^{\lambda_{12}+\omega_1-1}[<3]_r^{\lambda_3+\omega_2-\omega_1-1}\cdots[<\,n-2]_r^{\lambda_{n-2}+\omega_{n-3}-\omega_{n-4}-1}[n-1\,n]_r^{\lambda_{n-1}+\lambda_n-\omega_{n-3}-1}}{\Gamma(\lambda_{12}+\omega_1)\cdots\Gamma(\lambda_{n-1}+\lambda_n-\omega_{n-3})}\,.
\end{split}
\end{align}
By an appropriate change of variables dictated by 
\begin{equation}
    \lambda_{12}+\omega_1-1=k_1\,,\quad
    \lambda_3+\omega_2-\omega_1-1=k_2\,,\quad
    \cdots\quad
    \lambda_{n-1}+\lambda_n-\omega_{n-3}-1=k_{n-2}\,,
\end{equation}
and noting that we have
\begin{equation}
    \sum_{i=1}^{n-2}k_i=\sum_{i=1}^n\lambda_i-(n-2)\,.
\end{equation}
Using the multinomial formula, we find 
\begin{align}\label{npointdata}
    \begin{split}
    C_n(1,2,...,n)&=\frac{(rn)^{-2\lambda_n-2}}{\prod_{i=1}^{n-1}(ri)^{2\lambda_i-2}}\frac{(l_p)^{\Lambda_n-(n-2)}\alpha_n^{\Lambda_n-(n-2)}}{(\Lambda_n-(n-2))!}\,,\\
    \Lambda_n&=\sum_{i=1}^{n}\lambda_i\,,\quad
    \alpha_n\equiv \sum_{i<j}^{n-2}[ij]_r+[n-1\,n]_r=\sum_{i<j}^{n-1}[ij]_r\,.
    \end{split}
\end{align}
The total amplitude then becomes
\begin{align}
    \begin{split}
    \mathcal{A}_n^{\text{tree}}&=\frac{(rn)^{-2\lambda_n-2}}{\prod_{i=1}^{n-1}(ri)^{2\lambda_i-2}}\sum_{\sigma\in S_{n-1}}\frac{(l_p)^{\Lambda_n-(n-2)}\alpha_{n,\sigma}^{\Lambda_n-(n-2)}}{(\Lambda_n-(n-2))!}\tilde{\mathcal{A}}_n^{\text{SDYM}}(\sigma_1,\cdots, n)\mathrm{Tr}(T_{a_{\sigma_1}}\cdots T_{a_n})\,,
    \end{split}
\end{align}
where $\alpha_{n,\sigma}$ is the appropriate color-ordered version of $\alpha_{n}$. This is consistent with the light-cone result in \eqref{chiral_HISGRA_npt_LC}.

\paragraph{Example: (HS)-SDGR.} The relevant Poisson algebra structure constants cannot be represented by the trace structure used in \eqref{trace/commutator}, so the DDM basis is necessary. The DDM-ladder contains only kinematics and is 
\begin{align}\label{An_SDGR_SH}
    \begin{split}
    \mathbf{c}(1,2,...,n)&=\frac{(l_p)^{n-2}(rn)^2}{\prod_{i=1}^{n-1}(ri)^2}[12]_r[<3]_r\cdots [<n-2]_r[n-1,n]_r\\
    &=\frac{(l_p)^{n-2}(rn)^2}{\prod_{i=1}^{n-1}(ri)^2}\prod_{j=2}^{n-1}\left(\sum_{i<j}[ij]_r\right)\,.
    \end{split}
\end{align}
For HS-SDGR and numerous other theories \cite{Ponomarev:2017nrr,Krasnov:2021nsq,Serrani:2025owx} we find simply
\begin{align}\label{An_HS-SDGR_SH}
\mathbf{c}(1,2,...,n)=\mathcal{C}^{\lambda_1,\lambda_2,\cdots,\lambda_n}\frac{(rn)^{-2\lambda_n-2}}{\prod_{i=1}^{n-1}(ri)^{2\lambda_i-2}}\prod_{j=2}^{n-1}\left(\sum_{i<j}[ij]_r\right)\,.
\end{align}

\subsection{Generating functions: amplitudes for higher-spin multiplets}
Rather than considering amplitudes for fixed helicities, we would like to study the scattering of higher-spin multiplets. The natural way to do this is to sum over helicities and construct generating functions for amplitudes. This idea was already put forward in \cite{Ponomarev:2022atv} and resulted in quite peculiar amplitudes. The amplitudes we find below are essentially those of \cite{Ponomarev:2022atv} dressed with the $\text{sg}$-functions of \cite{Guevara:2026qwa}.

\paragraph{Example: Chiral HiSGRA.} 
We begin with the cubic amplitude in the form
\begin{align}
    \begin{split}
    \mathcal{A}_3&=\frac{l_p^{\Lambda_3-1}}{(\Lambda_3-1)!}\frac{(r1)^{\lambda_{23}-\lambda_1+1}(r2)^{\lambda_{13}-\lambda_2+1}}{(r3)^{2\lambda_3+2}}[12]^{\Lambda_3-1}\times \mathcal{A}^{\text{SDYM}}_3\,,\qquad \Lambda_n=\sum_{i=1}^n\lambda_i\,,\\
    &=\frac{l_p^{\Lambda_3-1}}{(\Lambda_3-1)!}\left(\frac{(r3)^2}{(r1)^2}\right)^{\lambda_1-1}\left(\frac{(r3)^2}{(r2)^2}\right)^{\lambda_2-1}\left(\frac{[12]_r}{(r3)^2}\right)^{\Lambda_3-1}\times \mathcal{A}^{\text{SDYM}}_3\,,
    \end{split}
\end{align}
where, in the second line, we have rewritten the expression in a form that will be convenient for the subsequent analysis. Summing over all helicities gives
\begin{align}
    \begin{split}
    \mathcal{A}^{\sum}_3&=\sum_{\Lambda_3=1}^{+\infty}\sum_{\lambda_{1,2}=-\infty}^{+\infty}\frac{l_p^{\Lambda_3-1}}{(\Lambda_3-1)!}\left(\frac{(r3)^2}{(r1)^2}\right)^{\lambda_1-1}\left(\frac{(r3)^2}{(r2)^2}\right)^{\lambda_2-1}\left(\frac{[12]_r}{(r3)^2}\right)^{\Lambda_3-1}\times \mathcal{A}^{\text{SDYM}}_3\,.\\
    &=e^{l_p\left(\frac{[12]_r}{(r3)^2}\right)}\delta\left(1-\frac{(r3)^2}{(r1)^2}\right)\delta\left(1-\frac{(r3)^2}{(r2)^2}\right)\times \mathcal{A}^{\text{SDYM}}_3\\
    &= \frac{(r3)^2}{(r1)(r2)} e^{l_p\left(\frac{[12]_r}{(r3)^2}\right)}\text{sg}[12] \,\delta (13)\delta (23)\delta\left(1-\frac{(r3)^2}{(r1)^2}\right)\delta\left(1-\frac{(r3)^2}{(r2)^2}\right)\hat{\delta}^{0|2}\\
    &=|(r1)(r2)|e^{l_p(\bar{1}\bar{2})}\text{sg}[12] \,\delta (13)\delta (23)\delta\left((r1)-(r3)\right)\delta\left((r2)-(r3)\right)\hat{\delta}^{0|2}\,.
    \end{split}
\end{align}
In the second line we have shifted the summation variables $\lambda_1$ and $\lambda_2$, and employed the regularization 
\begin{equation}    \sum_{\lambda\in\mathbb{Z}}z^{\lambda}\equiv\sum_{\lambda=-\infty}^{+\infty}z^{\lambda}=\delta(1-z)\,.
\end{equation}
This identity is standard in the vertex operator algebra (VOA) literature (see e.g. \cite{LepowskyLi2004,nozaradan2008introductionvertexalgebras}).The resulting structure may suggest a connection to an actual $2d$ dual CFT. In the third line, we have used
\begin{equation}\label{2terms}
    \delta\Big(1-\frac{(r3)^2}{(r1)^2}\Big)=(r1)^2\delta\Big((r1)^2-(r3)^2\Big)=\frac{|(r1)|}{2}\Big(\delta((r1)-(r3))+\delta((r1)+(r3))\Big)\,.
\end{equation}
Furthermore, if all external states are bosonic, the little-group transformation $\bar{i}_{A'}\rightarrow -\bar{i}_{A'}\,,\,i_A\rightarrow -i_A$ leaves the amplitude invariant
\begin{equation}
\mathcal{A}_n\rightarrow (-)^{2\lambda_i}\mathcal{A}_n=\mathcal{A}_n\,.
\end{equation}
Hence the generating field is symmetric $\Phi(i_A,\bar{i}_{A'})\rightarrow \Phi(-i_A,-\bar{i}_{A'})$. Therefore, the two terms in \eqref{2terms} act identically on bosonic test functions. Then, we can write
\begin{equation}
    \delta\Big(1-\frac{(r3)^2}{(r1)^2}\Big)=|(r1)|\delta((r1)-(r3))\,.
\end{equation}
Notice that by using the following identity 
\begin{equation}
     \delta^2(1_A-3_A)=\delta((\mu1)-(\mu3))\delta((r1)-(r3))|(\mu r)|=|(r1)|\delta((r1)-(r3))\delta(13)\,,
\end{equation}
the connection with the result of \cite{Ponomarev:2022atv} becomes manifest:
\begin{equation}\label{3ptgenF}
\mathcal{A}^{\sum}_3=\text{sg}[12]e^{l_p[12]}\delta^2(\bar{1}_{A'}+\bar{2}_{A'}+\bar{3}_{A'})\delta^2(1_A-2_A)\delta^2(1_A-3_A)\,.
\end{equation}
After stripping off the SDYM chamber function $A_{1,2,...,n}$ (in this case $\text{sg}[12]$), the remaining support and exponential kernel coincide with the ordered invariant trace of \cite{Ponomarev:2022atv}. 

Notice that a new effect, already described in \cite{Ponomarev:2022atv}, emerges: the summation over helicities generates two additional delta functions which, together with momentum conservation, give rise to the final two pairs of delta functions in \eqref{3ptgenF}. As a consequence, the holomorphic spinors become identical! As we shall see, this is a generic feature also for generating functions for higher-point amplitudes.

We could have started the analysis with the most general form of the cubic amplitude \eqref{general_cubic}, after factoring out the SDYM amplitude:
\begin{equation}
    \mathcal{A}_3=\frac{(r1)^{x+1}(r2)^{y+1}}{(r3)^{x+y+2}}[12]^{\lambda_{12}-\lambda_3+x+y-1}[23]^{\lambda_{23}-\lambda_1-x}[31]^{\lambda_{13}-\lambda_2-y}\times \mathcal{A}^{\text{SDYM}}_3\,,
\end{equation}
and summing would lead to 
\begin{equation}
    \mathcal{A}^{\sum_{\alpha,\beta}}_3=\frac{(r1)^{\alpha}(r2)^{\beta}}{(r3)^{\alpha+\beta}}\frac{[12]^{\alpha+\beta}}{[23]^{\alpha}[31]^{\beta}}\,\mathcal{A}^{\sum}_3\,,
\end{equation}
where we defined $\alpha=x+\lambda_1-\lambda_{23}$ and $\beta=y+\lambda_2-\lambda_{13}$. In fact, this prefactor is identically equal to one on the three-point support: the relations \eqref{momentum_conservation} give $[12]=\frac{(r3)}{(r1)}[23]=\frac{(r3)}{(r2)}[31]$, so that $\mathcal{A}^{\sum_{\alpha,\beta}}_3=\mathcal{A}^{\sum}_3$. The generating function is therefore independent of the representative chosen in \eqref{general_cubic}, as expected.

The same procedure can now be applied to the four-point amplitude, giving
\begin{align}\label{sum_s_U(n)_chiral}
    \begin{split}
    \mathcal{A}^{\sum}_s&=\sum_{\Lambda_4=2}^{+\infty}\sum_{\lambda_{1,2,3}=-\infty}^{+\infty}\frac{l_p^{\Lambda_4-2}}{(\Lambda_4-2)!}\left(\frac{(r4)^2}{(r1)^2}\right)^{\lambda_1-1}\left(\frac{(r4)^2}{(r2)^2}\right)^{\lambda_2-1}\left(\frac{(r4)^2}{(r3)^2}\right)^{\lambda_3-1}\times\\
    &\quad\left(\frac{[12]_r+[34]_r}{(r4)^2}\right)^{\Lambda_4-2}\times \mathcal{A}^{\text{SDYM}}_s\\
    &=e^{l_p\left(\frac{[12]_r+[34]_r}{(r4)^2}\right)}\delta\left(1-\frac{(r4)^2}{(r1)^2}\right)\delta\left(1-\frac{(r4)^2}{(r2)^2}\right)\delta\left(1-\frac{(r4)^2}{(r3)^2}\right)\times \mathcal{A}^{\text{SDYM}}_s\\
    &=\left|\prod_{i=1}^{3}(ri)\right|e^{l_p([12]+[34])}\text{sg}[12]\text{sg}[34] \prod_{a=1}^{3}\delta (a4)\prod_{j=1}^{3}\delta\left((rj)-(r4)\right)\hat{\delta}^{0|2}\\
    &=\text{sg}[12]\text{sg}[34]e^{l_p([12]+[34])}\delta^2\left(\sum_{i=1}^4\bar{i}_{A'}\right)\prod_{j=1}^{3}\delta^2(j_A-4_A)\,.
    \end{split}
\end{align}
Again, apart from the overall sign factors, this expression coincides exactly with Eq. $(7.2)$ of \cite{Ponomarev:2022atv}. Indeed, on the support of the delta functions
\begin{equation}
    [12]+[34]=[23]+[24]+[34]\,,
\end{equation}
in the delta function support of \eqref{sum_s_U(n)_chiral}.
An analogous expression is obtained in the $t$- and $u$-channels. For example,
\begin{equation}
    \mathcal{A}^{\sum}_t=\text{sg}[41]\text{sg}[23]e^{l_p([41]+[23])}\delta^2\left(\sum_{i=1}^4\bar{i}_{A'}\right)\prod_{j=1}^{3}\delta^2(j_A-4_A)\,.
\end{equation}
The generalization to arbitrary multiplicity is straightforward:
\begin{align}
    \begin{split}
    \mathcal{A}^{\sum}_n&=\sum_{\Lambda_n=n-2}^{+\infty}\sum_{\lambda_i=-\infty}^{+\infty}\frac{l_p^{\Lambda_n-(n-2)}\alpha^{\Lambda_n-(n-2)}_n}{(\Lambda_n-(n-2))!}\times\frac{(rn)^{-2\lambda_n-2}}{\prod_{i=1}^{n-1}(ri)^{2\lambda_i-2}}\times\mathcal{A}^{\text{SDYM}}_n\\
    &=e^{l_p\left(\frac{\alpha_n}{(rn)^2}\right)}\prod_{i=1}^{n-1}\delta\left(1-\frac{(rn)^2}{(ri)^2}\right)\times \mathcal{A}^{\text{SDYM}}_n\\
    &=\left|\prod_{i=1}^{n-1}(ri)\right|
    A_{1,2,..,n}e^{l_p\left(\frac{\alpha_n}{(rn)^2}\right)}\prod_{a=1}^{n-1}\delta (an)\prod_{j=1}^{n-1}\delta\left((rj)-(rn)\right)\hat{\delta}^{0|2}\\
    &=A_{1,2,..,n}e^{l_p\left(\frac{\alpha_n}{(rn)^2}\right)}\delta^2\left(\sum_{i=1}^n\bar{i}_{A'}\right)\prod_{j=1}^{n-1}\delta^2(j_A-n_A)\,,
    \end{split}
\end{align}
where the fraction in the exponent is just ${\alpha_n}/{(rn)^2}=\sum_{i<j}^{n-1}[ij]$. This is one color-ordered partial amplitude. The full Chiral HiSGRA amplitude requires the sum over $\sigma\in S_{n-1}$.
Apart from the multiplicative SDYM chamber function $A_{1,2,..,n}$, this expression agrees exactly with Eq. $(7.1)$ of \cite{Ponomarev:2022atv}. This follows after using momentum conservation
\begin{equation}
    \sum_{i<j}^{n-2}[ij]+[n-1,n]=\sum_{2\leq i< j \leq n}[ij]\,.
\end{equation}
This can be easily verified using
\begin{equation}\label{useful_1}
    \sum_{i=1}^n\bar{i}_{A'}=0
    \implies
    [n-1,n]=\sum_{i< n-1}[i,n-1]\,,\quad
    [nn]=0=\sum_{j< n} [jn]\,,\quad
    \sum_{i=1}^n [1i]=0\,.
    \end{equation}
Using the first identity we see that 
\begin{equation}
    \sum_{i<j}^{n-2}[ij]+[n-1,n]=\sum_{i<j\leq n-1}[ij]\,,
\end{equation}
and via the other two, it is simple to verify
\begin{equation}
    \sum_{i< j\leq n-1}[ij]-\sum_{2\leq i< j \leq n}[ij]=[n1]-[n1]=0\,.
\end{equation}
The same analysis can now be carried out for the HS-SDYM and HS-SDGR theories.

\paragraph{Example: HS-SDYM. } Here $\Lambda_n=n-2$.
\begin{align}
    \begin{split}
    \mathcal{A}^{\text{HS-SDYM},\,\sum}_n&=\sum_{\lambda_i=-\infty}^{+\infty}g^{n-2}\frac{(rn)^{-2\lambda_n-2}}{\prod_{i=1}^{n-1}(ri)^{2\lambda_i-2}}\times\mathcal{A}^{\text{SDYM}}_n\\
    &=g^{n-2}\left|\prod_{i=1}^{n-1}(ri)\right|
    A_{1,2,..,n}\prod_{a=1}^{n-1}\delta (an)\prod_{j=1}^{n-1}\delta\left((rj)-(rn)\right)\hat{\delta}^{0|2}\\
    &=g^{n-2}A_{1,2,..,n}\delta^2\left(\sum_{i=1}^n\bar{i}_{A'}\right)\prod_{j=1}^{n-1}\delta^2(j_A-n_A)\,.
    \end{split}
\end{align}
This is one color-ordered partial amplitude. The full HS-SDYM amplitude requires the sum over $\sigma\in S_{n-1}$.

\paragraph{Example: HS-SDGR.} Here $\Lambda_n=2(n-2)$.
\begin{align}
    \begin{split}
    \mathcal{A}&^{\text{HS-SDGR},\,\sum}_n=\sum_{\lambda_i=-\infty}^{+\infty}(l_p)^{n-2}\prod_{j=2}^{n-1}[<j]_r\times\frac{(rn)^{-2\lambda_n-2}}{\prod_{i=1}^{n-1}(ri)^{2\lambda_i-2}}\times\mathcal{A}^{\text{SDYM}}_n\\
&=(l_p)^{n-2}\left|\prod_{i=1}^{n-1}(ri)\right|
    A_{1,2,..,n}\prod_{j=2}^{n-1}\frac{[<j]_r}{(rn)^2}\prod_{a=1}^{n-1}\delta (an)\prod_{k=1}^{n-1}\delta\left((rk)-(rn)\right)\hat{\delta}^{0|2}\\
    &=(l_p)^{n-2}A_{1,2,..,n}\prod_{j=2}^{n-1}[<j]\delta^2\left(\sum_{i=1}^n\bar{i}_{A'}\right)\prod_{k=1}^{n-1}\delta^2(k_A-n_A)\,.
    \end{split}
\end{align}
This is one specific DDM-ordered contribution. The full HS-SDGR amplitude still requires the sum over $\sigma\in S_{n-2}$.

To conclude, we have shown that the spinor-helicity amplitudes of any SD-theory factor into the SDYM chamber function $A_{1,2,...,n}$ and a theory-dependent DDM dressing, fixed entirely by the specific kinematical algebra. In Chiral HiSGRA, the formal helicity sum reproduces the support and exponential kernel of the ordered flat-space higher-spin invariants of \cite{Ponomarev:2022atv}, while retaining the prescription-dependent SDYM chamber function. 

\section{\label{sec:conclusions}Conclusions \& Discussion}
In this paper we developed a unified treatment of SD-theories, both in the light-cone gauge and in the spinor-helicity formalism. The key input is the statement that behind any SD-theory there is a kinematic Lie algebra, \cite{Monteiro:2011pc, Ponomarev:2017nrr,Serrani:2025oaw}. While this is a very deep statement, it may not be obvious how to apply it to concrete problems in field theory. At least for tree-level diagrams, the DDM-decomposition reduces amplitudes in any SD-theory to the sum over partial SDYM amplitudes times the DDM ``color'' ladders built now with structure constants of a kinematic algebra. Basically, the DDM-ordering allows us to formulate a ``double copy'' procedure that is different from the well-known KLT \cite{Kawai:1985xq}, BCJ \cite{Bern:2010ue}, or CHY \cite{Cachazo:2013iea}, but is conceptually close to BCJ. The main difference is that a given kinematic algebra may not define a theory on its own, it merely leads to a SD-theory when used together with the kinematic vertex of SDYM.\footnote{There are many other SD-theories that can be obtained as double- and multi-copy of the basic theories discussed in the present paper \cite{Ponomarev:2024jyg}. Our conclusions should apply to those theories as well. }

An immediate consequence is that all SD-theories have nonvanishing amplitudes (in complex kinematics or in the split signature) thanks to \cite{Guevara:2026qzd}, see also \cite{Guevara:2026qwa,Brandhuber:2026njb}. In particular, since most SD-theories contain higher-spin fields, there is a large class of higher-spin theories with nonvanishing amplitudes in flat space, including the maximal SD-theory --- Chiral HiSGRA. Another interesting consequence is that amplitudes in all SD-theories reduce to those of SDYM, which may sound counterintuitive for non-colored theories like SDGR. Lastly, the DDM-decomposition is well-established at one-loop, and hence, our main conclusions can be applied to one-loop amplitudes as well, which, for example, explains why one-loop amplitudes of Chiral HiSGRA \cite{Skvortsov:2020gpn} reduce to those of SDYM.

One significant improvement as compared to an earlier version of the paper is Section \ref{sec:spinorhell}, where $n$-point amplitudes of SD-theories were obtained within the manifestly Lorentz-invariant spinor-helicity formalism. This required an off-shell extension of the on-shell three-point vertices (amplitudes), for which we used the CSW formalism. This way the $n$-point amplitudes can be compared with the invariants computed in \cite{Ponomarev:2022atv}.   

Let us note that many ``purely'' higher-spin results have immediate low-spin consequences. As it was already noted, tree-level amplitudes of SDYM and SDGR are subsets of those of the Chiral HiSGRA (we consider the maximal SD-theory, for simplicity). The same is true for the one-loop amplitudes of SDYM and SDGR. For example, the one-loop amplitude of the $U(N)$-gauged Chiral HiSGRA has the form of the all-helicity-plus one-loop amplitude of SDYM times a certain kinematic factor \cite{Skvortsov:2020gpn}. 

Since Chiral HiSGRA has extended symmetries, which are visible in a covariant formulation constructed in \cite{Sharapov:2022faa,Sharapov:2022wpz},\footnote{In the manifestly gauge-invariant and Lorentz-covariant formulation, there is an infinite-dimensional algebra of gauge symmetries, which is $A_1\otimes \mathbb{C}[y^A]$, i.e. it is a tensor product of the Weyl algebra $A_1$ and a commutative algebra. The Lie algebra of symmetries is the commutator Lie algebra of this associative algebra. Upon reduction to the light-cone gauge, the gauge symmetry algebra is converted to the kinematic algebra.} one can envisage a few consequences for low-spin theories. Invariants of the higher-spin algebra are the lowest approximation to higher-spin invariant observables.\footnote{The idea of higher-spin invariant observables goes back to \cite{Sezgin:2011hq}. It was in \cite{Colombo:2012jx,Didenko:2012tv} that the invariants of a particular higher-spin algebra were computed and gave all $n$-point functions of higher-spin currents in free vector models, see also \cite{Bonezzi:2017vha}.} In the AdS-setting the simplest such observables take the form $\mathcal{O}_n=\mathrm{Tr}[A_1\star A_2\star ...\star A_n]$ with $\star$ being the product in the (associative) higher-spin algebra of interest. 

It was shown in \cite{Didenko:2012tv} that $\mathcal{O}_n$ computes $n$-point correlators of conserved higher-spin currents in free vector models.\footnote{We should stress that one needs to ``know'' what to compute the invariants on, i.e. what to plug into the traces. The ``wave-functions'' used in \cite{Didenko:2012tv} were adapted to the free vector model/HS duality \cite{Klebanov:2002ja, Sezgin:2002rt, Leigh:2003gk}.} Chiral HiSGRA is smooth in the cosmological constant, and there is a direct analog of $\mathcal{O}_n$ in flat space, which was introduced in \cite{Ponomarev:2022atv}. For the flat space higher-spin algebra, $\mathcal{O}_n$ are still nontrivial, of distributional nature and display collinear features. The missing piece of the puzzle has been that it was not clear how the nontriviality of the higher-spin invariants is consistent with vanishing amplitudes in Chiral HiSGRA. This has been just resolved for $(+...+-)$ amplitudes in Yang-Mills theory \cite{Guevara:2026qzd} and, as we show, has immediate consequences for all SD-theories, including those with higher-spin fields.

The same invariants $\mathcal{O}_n$ are equal to the correlators of the ``flat space singleton'' \cite{Ponomarev:2022qkx,Ponomarev:2022ryp}, which is an analog of the free vector model but for flat holography. This is not too surprising since the amplitudes and the correlators are fixed by the same infinite-dimensional algebra. Now, the amplitudes of Chiral HiSGRA are shown to be nontrivial. What needs to be done is to demonstrate that there is a scheme where the amplitudes of Chiral HiSGRA are given by the same invariants. It may not be straightforward since the invariants need to be computed on concrete ``wave-functions'' that encode the nature of the Green's functions being used to compute amplitudes. An encouraging observation \cite{Ponomarev:2022atv} is that at the three-point level, the amplitudes of Chiral HiSGRA do have collinear features. 

Therefore, within celestial holography, see e.g. \cite{Strominger:2017zoo} for a review, there should exist a direct analog of the higher-spin/vector-model AdS/CFT duality \cite{Klebanov:2002ja,Sezgin:2002rt,Leigh:2003gk} that is between the flat space singleton and Chiral HiSGRA as was proposed by Ponomarev in \cite{Ponomarev:2022qkx,Ponomarev:2022ryp,Ponomarev:2022atv}. This duality is not obstructed by the nonlocality contrary to the duals of vector models \cite{Bekaert:2010hp,Bekaert:2015tva,Maldacena:2015iua,Sleight:2017pcz,Ponomarev:2017qab}. A version of the vector-model/HS AdS/CFT duality that is free of the nonlocality problem is between Chiral HiSGRA and a subsector of Chern--Simons vector-models \cite{Sharapov:2022awp,Aharony:2024nqs,Jain:2024bza}. It would also be interesting to find out to what extent the celestial duality is a limit of the AdS/CFT one.

One low-spin consequence of higher-spin symmetries is that amplitudes in SDYM and SDGR can also be computed via the higher-spin invariant observables $\mathcal{O}_n$ and just correspond to the lowest orders in the spin expansion. Note, however, that SDYM/SDGR amplitudes are not themselves higher-spin invariant, rather they are particular components of a higher-spin invariant quantity. This has to be true both in flat and (A)dS spaces. It is known that the leading energy-pole of AdS/CFT correlators is equal to the flat space amplitude in the same theory \cite{Maldacena:2011nz}. Now that such amplitudes are nontrivial in SD-theories, it would be interesting to investigate the CFT consequences. 

Following \cite{Ponomarev:2017nrr}, one can conclude that all SD-theories are equivalent to the principal chiral model, with the kinematic algebra being the underlying symmetry. In particular, all SD-theories are integrable. In the covariant formulation of Chiral HiSGRA \cite{Sharapov:2022wpz}, which extends it to nonvanishing cosmological constant, the $2d$-origin is also visible. Therefore, a stringy reformulation of Chiral HiSGRA is plausible. 
 
Lastly, the nontriviality of collinear amplitudes can give more room for higher-spin theories to have interesting $S$-matrices. One can also imagine a parity-invariant completion of Chiral HiSGRA where particles can scatter both via holomorphic and anti-holomorphic collinear processes. 

\section*{Acknowledgments}
E.S. is grateful to Thomas Basile, Chanon Hasuwannakit, Kirill Krasnov, Alexander Ochirov, Mitya Ponomarev, David Skinner, Alexey Sharapov, Stephan Stieberger, and Andrew Strominger for useful discussions. We are grateful to Mitya Ponomarev for useful comments on the draft of this paper. E.S. is also grateful to the Simons Collaboration on Celestial Holography for support during the last stage of this work. The work of M.S. and E.S. was partially supported by the European Research Council (ERC) under the European Union’s Horizon 2020 research and innovation programme (grant agreement No 101002551). E.S. is a research associate of the Fonds de la Recherche Scientifique – FNRS. 

\appendix 

\section{Light-cone}\label{app:LC}
In the light-cone Hamiltonian approach for massless higher-spin fields, we have 
\begin{equation}\label{diag_momenta}
    p_{iAA'}=p_{i\mu}(\sigma^{\mu})_{AA'}=\sqrt{2}
    \begin{pmatrix}
        p_i^- & p_i\\
        \bar{p}_i & - \beta_i
    \end{pmatrix}\approx\sqrt{2}
    \begin{pmatrix}
        -\frac{p_i\bar{p}_i}{\beta_i} & p_i\\
        \bar{p}_i & - \beta_i
    \end{pmatrix}=i_A\bar{i}_{A'}\,.
\end{equation}
Here $\approx$ denotes the mass-shell condition
$p_i^-=-p_i\bar p_i/\beta_i$, under which $p_{iAA'}$ factorizes as
$i_A\bar i_{A'}$. Notice that for complex momenta $p_i^{\mu}$ the two spinors $(i_A,\bar{i}_{A'})$ are independent two-dimensional complex vectors. In Minkowski space and for real momenta $p_{iAA'}$ is hermitian, and the two spinors become complex conjugate $\bar{i}_{A'}=\pm(i_A)^*$ (where the sign depends on whether the energy is taken to be positive or negative, then on the convention we use on the background flat metric).

Our spinor-helicity convention for massless particles is
\begin{align}\label{SH_amplitude_def}
    &i_A=\sqrt{2}(z_i,1)\,,&
    &\bar{i}_{A'}=-\omega_i(\bar{z}_i,1)\,,&
    &(ij)=-\epsilon^{AB}i_Aj_B\,,&
&(\bar{i}\bar{j})=-\epsilon^{A'B'}\,\bar{i}_{A'}\bar{j}_{B'}\,,
\end{align}
where
\begin{align}
    &\epsilon^{AB}=\epsilon_{AB}=
    \begin{pmatrix}
        0 & 1\\
        -1 & 0
    \end{pmatrix}\,,&
    &\epsilon^{A'B'}=\epsilon_{A'B'}=
    \begin{pmatrix}
        0 & 1\\
        -1 & 0
    \end{pmatrix}\,.
\end{align}
We raise and lower spinor indices according to the convention $i^A=\epsilon^{AB}i_B$ and $\epsilon_{AB}i^B=-i_A$. By choosing the spinors that diagonalize the momenta and respect \eqref{diag_momenta}, we get
\begin{align}
&\omega_i=\beta_i\,,&
&z_i=-\frac{p_i}{\beta_i}\,,&
&\bar{z}_i=-\frac{\bar{p}_i}{\beta_i}&
&\implies&
&i_A=\frac{\sqrt{2}}{\beta_i}(-p_i,\beta_i)\,,&
&\bar{i}_{A'}=(\bar{p}_i,-\beta_i)\,.
\end{align}
Therefore, we have
\begin{align}\label{Pbeta_norm}
\begin{aligned}
        (\bar{i}\bar{j})&=-\omega_i\omega_j\bar{z}_{ij}=\PPb_{ij}\,,\qquad && (ij)=-2z_{ij}=\frac{2}{\beta_i\beta_j}\PP_{ij}\equiv \PP^{\beta}_{ij}\,, \\
    \PPb_{ij}&=\bar{p}_i\beta_j-\bar{p}_j\beta_i\,,
&&\PP_{ij}=p_i\beta_j-p_j\beta_i\,,
\end{aligned}
\end{align}
where $z_{ij}=z_i-z_j$ and $\bar{z}_{ij}=\bar{z}_i-\bar{z}_j$. Spinor-helicity variables $(i_A,\bar{i}_{A'})$ are defined up to little group scaling $(i_A,\bar{i}_{A'})\sim (t_ii_A,t_i^{-1}\bar{i}_{A'})$ for $t_i\in\mathbb{C}\setminus\{0\}$.

\section{Useful relations}\label{app:identities}
Some on-shell relations for the $n$-point scattering are
\begin{align}
   &-\frac{2}{\beta_i\beta_j}\PP_{ij}\PPb_{ij}=2\,p_i \cdot p_j=(p_i+p_j)^2\,,&
   &\sum^{n}_{j=1}p^{\mu}_j=0\quad\implies\quad\sum^{n}_{j=1}\frac{\PP_{ij}\PPb_{jk}}{\beta_j}=0\,.
\end{align}
In particular, for the four-point scattering, we find
\begin{align}
    &\frac{\PPb_{12}\PPb_{34}}{(p_1+p_2)^2}=\frac{\PPb_{31}\PPb_{24}}{(p_1+p_3)^2}=\frac{\PPb_{14}\PPb_{23}}{(p_1+p_4)^2}\,,&
    &s_{ij}=-(p_i+p_j)^2\,,
\end{align}
where $s=s_{12}$, $t=s_{14}$, $u=s_{13}$ are the standard Mandelstam variables for massless four-point scattering. Off-shell, we have the following relations for the $n$-point scattering:
\begin{align}\label{offshell1}
    &\sum_j\frac{\PP_{ij}\PPb_{jk}}{\beta_j}=-\frac{\beta_i\beta_k}{2}\sum_j\frac{p_j^2}{\beta_j}\,,\\\label{offschell2}
    &\PP_{ik}\PPb_{ik}=\frac{\beta_i\beta_k}{2}s_{ik}+\frac{\beta_i}{2}(\beta_i+\beta_k)p_k^2\,,\quad p_i^2=0\,,\quad p_k^2\neq 0\,.
\end{align}
In two dimensions, for $(\mu r)\neq 0$, we can always decompose any two-component vector as 
\begin{equation}\label{SHproperty1}
    i_A=\frac{(\mu i)}{(\mu r)}r_A+\frac{(ir)}{(\mu r)}\mu_A\,.
\end{equation} 
When two spinors are collinear (i.e. $i_A\parallel j_A$) we have 
\begin{equation}
    i_A=\frac{(ri)}{(rj)}j_A\,.
\end{equation}
The momentum conservation can be massaged as
\begin{align}
    \begin{split}
    &\delta^4\left(\sum_i i_A\bar{i}_{A'}\right)= \delta^4\left(\frac{1}{(\mu r)}\sum_i((\mu i)r_A+(ri)\mu_A)\bar{i}_{A'}\right)=(\mu r)^4\delta^4\left(\sum_i((\mu i)r_A+(ri)\mu_A)\bar{i}_{A'}\right)\\
    &=(\mu r)^2\delta^2\left(\sum_i(\mu i)\bar{i}_{A'}\right)\delta^2\left(\sum_i(ri)\bar{i}_{A'}\right)\equiv (\PP^{\beta}_{\mu r})^2\delta^2\left(\sum_i\PP^{\beta}_{\mu i}\bar{i}_{A'}\right)\delta^2\left(\sum_i\PP^{\beta}_{ri}\bar{i}_{A'}\right)\,,
    \end{split}
\end{align}
where the first equality follows from \eqref{SHproperty1}, assuming $(\mu r) \neq 0$, and the last equality results from a change of variables. Momentum conservation then takes the form
Another useful relation is
\begin{equation}
    \delta^2(a\,i_A+b\,j_A)=\frac{\delta(a)\delta(b)}{|(ij)|}\,.
\end{equation}

\footnotesize
\providecommand{\href}[2]{#2}\begingroup\raggedright\endgroup


\begin{thebibliography}{100}

\bibitem{Ponomarev:2016lrm}
D.~Ponomarev and E.~D. Skvortsov, ``{Light-Front Higher-Spin Theories in Flat Space},'' {\em J. Phys.} {\bf A50} (2017), no.~9, 095401,
\href{http://arXiv.org/abs/1609.04655}{{\tt 1609.04655}}.

\bibitem{Ponomarev:2017nrr}
D.~Ponomarev, ``{Chiral Higher Spin Theories and Self-Duality},'' {\em JHEP} {\bf 12} (2017) 141,
\href{http://arXiv.org/abs/1710.00270}{{\tt 1710.00270}}.

\bibitem{Krasnov:2021nsq}
K.~Krasnov, E.~Skvortsov, and T.~Tran, ``{Actions for self-dual Higher Spin Gravities},'' {\em JHEP} {\bf 08} (2021) 076, \href{http://arXiv.org/abs/2105.12782}{{\tt 2105.12782}}.

\bibitem{Monteiro:2022xwq}
R.~Monteiro, ``{From Moyal deformations to chiral higher-spin theories and to celestial algebras},'' {\em JHEP} {\bf 03} (2023) 062, \href{http://arXiv.org/abs/2212.11266}{{\tt 2212.11266}}.

\bibitem{Ponomarev:2024jyg}
D.~Ponomarev, ``{Chiral higher-spin double copy},'' {\em JHEP} {\bf 01} (2025) 143, \href{http://arXiv.org/abs/2409.19449}{{\tt 2409.19449}}.

\bibitem{Serrani:2025owx}
M.~Serrani, ``{On classification of (self-dual) higher-spin gravities in flat space},'' {\em JHEP} {\bf 08} (2025) 032, \href{http://arXiv.org/abs/2505.12839}{{\tt 2505.12839}}.

\bibitem{Fronsdal:1978rb}
C.~Fronsdal, ``Massless fields with integer spin,'' {\em Phys. Rev.} {\bf D18} (1978)
3624.

\bibitem{Fradkin:1986ka}
E.~S. Fradkin and M.~A. Vasiliev, ``Candidate to the role of higher spin symmetry,'' {\em Ann. Phys.} {\bf 177} (1987)
63.

\bibitem{Flato:1978qz}
M.~Flato and C.~Fronsdal, ``{One Massless Particle Equals Two Dirac Singletons: Elementary Particles in a Curved Space. 6.},'' {\em Lett.Math.Phys.} {\bf 2} (1978)
421--426.

\bibitem{Berends:1984wp}
F.~A. Berends, G.~J.~H. Burgers, and H.~Van~Dam, ``{On spin three selfinteractions},'' {\em Z. Phys.} {\bf C24} (1984)
247--254.

\bibitem{Maldacena:2011jn}
J.~Maldacena and A.~Zhiboedov, ``{Constraining Conformal Field Theories with A Higher Spin Symmetry},''
\href{http://arXiv.org/abs/1112.1016}{{\tt 1112.1016}}.

\bibitem{Boulanger:2013zza}
N.~Boulanger, D.~Ponomarev, E.~D. Skvortsov, and M.~Taronna, ``{On the uniqueness of higher-spin symmetries in AdS and CFT},'' {\em Int. J. Mod. Phys.} {\bf A28} (2013) 1350162,
\href{http://arXiv.org/abs/1305.5180}{{\tt 1305.5180}}.

\bibitem{Blencowe:1988gj}
M.~Blencowe, ``{A Consistent Interacting Massless Higher Spin Field Theory in $D$ = (2+1)},'' {\em Class.Quant.Grav.} {\bf 6} (1989)
443.

\bibitem{Bergshoeff:1989ns}
E.~Bergshoeff, M.~P. Blencowe, and K.~S. Stelle, ``{Area Preserving Diffeomorphisms and Higher Spin Algebra},'' {\em Commun. Math. Phys.} {\bf 128} (1990)
213.

\bibitem{Campoleoni:2010zq}
A.~Campoleoni, S.~Fredenhagen, S.~Pfenninger, and S.~Theisen, ``{Asymptotic symmetries of three-dimensional gravity coupled to higher-spin fields},'' {\em JHEP} {\bf 1011} (2010) 007,
\href{http://arXiv.org/abs/1008.4744}{{\tt 1008.4744}}.

\bibitem{Henneaux:2010xg}
M.~Henneaux and S.-J. Rey, ``{Nonlinear $W_{\infty}$ as Asymptotic Symmetry of Three-Dimensional Higher Spin Anti-de Sitter Gravity},'' {\em JHEP} {\bf 1012} (2010) 007,
\href{http://arXiv.org/abs/1008.4579}{{\tt 1008.4579}}.

\bibitem{Grigoriev:2020lzu}
M.~Grigoriev, K.~Mkrtchyan, and E.~Skvortsov, ``{Matter-free higher spin gravities in 3D: Partially-massless fields and general structure},'' {\em Phys. Rev. D} {\bf 102} (2020), no.~6, 066003, \href{http://arXiv.org/abs/2005.05931}{{\tt 2005.05931}}.

\bibitem{Pope:1989vj}
C.~N. Pope and P.~K. Townsend, ``{Conformal Higher Spin in (2+1)-dimensions},'' {\em Phys. Lett.} {\bf B225} (1989)
245--250.

\bibitem{Fradkin:1989xt}
E.~S. Fradkin and V.~{\relax Ya}. Linetsky, ``{A Superconformal Theory of Massless Higher Spin Fields in $D$ = (2+1)},'' {\em Mod. Phys. Lett.} {\bf A4} (1989) 731.
[Annals Phys.198,293(1990)].

\bibitem{Grigoriev:2019xmp}
M.~Grigoriev, I.~Lovrekovic, and E.~Skvortsov, ``{New Conformal Higher Spin Gravities in $3d$},'' {\em JHEP} {\bf 01} (2020) 059,
\href{http://arXiv.org/abs/1909.13305}{{\tt 1909.13305}}.

\bibitem{Basile:2024raj}
T.~Basile, ``{Massless chiral fields in six dimensions},'' {\em SciPost Phys.} {\bf 19} (2025), no.~3, 079, \href{http://arXiv.org/abs/2409.12800}{{\tt 2409.12800}}.

\bibitem{Skvortsov:2020wtf}
E.~Skvortsov, T.~Tran, and M.~Tsulaia, ``{More on Quantum Chiral Higher Spin Gravity},'' {\em Phys. Rev.} {\bf D101} (2020), no.~10, 106001,
\href{http://arXiv.org/abs/2002.08487}{{\tt 2002.08487}}.

\bibitem{Skvortsov:2018jea}
E.~D. Skvortsov, T.~Tran, and M.~Tsulaia, ``{Quantum Chiral Higher Spin Gravity},'' {\em Phys. Rev. Lett.} {\bf 121} (2018), no.~3, 031601,
\href{http://arXiv.org/abs/1805.00048}{{\tt 1805.00048}}.

\bibitem{Skvortsov:2020gpn}
E.~Skvortsov and T.~Tran, ``{One-loop Finiteness of Chiral Higher Spin Gravity},'' {\em JHEP} {\bf 07} (2020) 021, \href{http://arXiv.org/abs/2004.10797}{{\tt 2004.10797}}.

\bibitem{Tsulaia:2022csz}
M.~Tsulaia and D.~Weissman, ``{Supersymmetric quantum chiral higher spin gravity},'' {\em JHEP} {\bf 12} (2022) 002, \href{http://arXiv.org/abs/2209.13907}{{\tt 2209.13907}}.

\bibitem{Ponomarev:2022atv}
D.~Ponomarev, ``{Invariant traces of the flat space chiral higher-spin algebra as scattering amplitudes},'' {\em JHEP} {\bf 09} (2022) 086, \href{http://arXiv.org/abs/2205.09654}{{\tt 2205.09654}}.

\bibitem{Bekaert:2010hp}
X.~Bekaert, N.~Boulanger, and S.~Leclercq, ``{Strong obstruction of the Berends-Burgers-van Dam spin-3 vertex},'' {\em J.Phys.} {\bf A43} (2010) 185401,
\href{http://arXiv.org/abs/1002.0289}{{\tt 1002.0289}}.

\bibitem{Roiban:2017iqg}
R.~Roiban and A.~A. Tseytlin, ``{On four-point interactions in massless higher spin theory in flat space},'' {\em JHEP} {\bf 04} (2017) 139,
\href{http://arXiv.org/abs/1701.05773}{{\tt 1701.05773}}.

\bibitem{Serrani:2026dbs}
M.~Serrani, ``{Massless spinning fields on the Light-Front: quartic vertices and amplitudes},'' \href{http://arXiv.org/abs/2602.12826}{{\tt 2602.12826}}.

\bibitem{Dempster:2012vw}
P.~Dempster and M.~Tsulaia, ``{On the Structure of Quartic Vertices for Massless Higher Spin Fields on Minkowski Background},'' {\em Nucl. Phys.} {\bf B865} (2012) 353--375,
\href{http://arXiv.org/abs/1203.5597}{{\tt 1203.5597}}.

\bibitem{Bekaert:2015tva}
X.~Bekaert, J.~Erdmenger, D.~Ponomarev, and C.~Sleight, ``{Quartic AdS Interactions in Higher-Spin Gravity from Conformal Field Theory},'' {\em JHEP} {\bf 11} (2015) 149,
\href{http://arXiv.org/abs/1508.04292}{{\tt 1508.04292}}.

\bibitem{Maldacena:2015iua}
J.~Maldacena, D.~Simmons-Duffin, and A.~Zhiboedov, ``{Looking for a bulk point},'' {\em JHEP} {\bf 01} (2017) 013,
\href{http://arXiv.org/abs/1509.03612}{{\tt 1509.03612}}.

\bibitem{Sleight:2017pcz}
C.~Sleight and M.~Taronna, ``{Higher-Spin Gauge Theories and Bulk Locality},'' {\em Phys. Rev. Lett.} {\bf 121} (2018), no.~17, 171604,
\href{http://arXiv.org/abs/1704.07859}{{\tt 1704.07859}}.

\bibitem{Ponomarev:2017qab}
D.~Ponomarev, ``{A Note on (Non)-Locality in Holographic Higher Spin Theories},'' {\em Universe} {\bf 4} (2018), no.~1, 2,
\href{http://arXiv.org/abs/1710.00403}{{\tt 1710.00403}}.

\bibitem{Sharapov:2024euk}
A.~Sharapov, E.~Skvortsov, and A.~Sukhanov, ``{Matter-coupled higher spin gravities in 3d: no- and yes-go results},'' {\em JHEP} {\bf 04} (2025) 155, \href{http://arXiv.org/abs/2409.12830}{{\tt 2409.12830}}.

\bibitem{Bekaert:2025azj}
X.~Bekaert, A.~Sharapov, and E.~Skvortsov, ``{Higher-spin Poisson sigma models and holographic duality for SYK models},'' {\em JHEP} {\bf 04} (2026) 063, \href{http://arXiv.org/abs/2509.19964}{{\tt 2509.19964}}.

\bibitem{Alkalaev:2013fsa}
K.~Alkalaev, ``{On higher spin extension of the Jackiw-Teitelboim gravity model},'' {\em J.Phys.} {\bf A47} (2014) 365401,
\href{http://arXiv.org/abs/1311.5119}{{\tt 1311.5119}}.

\bibitem{Alkalaev:2020kut}
K.~Alkalaev and X.~Bekaert, ``{On BF-type higher-spin actions in two dimensions},'' {\em JHEP} {\bf 05} (2020) 158, \href{http://arXiv.org/abs/2002.02387}{{\tt 2002.02387}}.

\bibitem{Alkalaev:2019xuv}
K.~Alkalaev and X.~Bekaert, ``{Towards higher-spin AdS$_2$/CFT$_1$ holography},'' {\em JHEP} {\bf 04} (2020) 206, \href{http://arXiv.org/abs/1911.13212}{{\tt 1911.13212}}.

\bibitem{Metsaev:1991mt}
R.~R. Metsaev, ``{Poincare invariant dynamics of massless higher spins: Fourth order analysis on mass shell},'' {\em Mod. Phys. Lett.} {\bf A6} (1991)
359--367.

\bibitem{Metsaev:1991nb}
R.~R. Metsaev, ``{$S$ matrix approach to massless higher spins theory. 2: The Case of internal symmetry},'' {\em Mod. Phys. Lett.} {\bf A6} (1991)
2411--2421.

\bibitem{Serrani:2025oaw}
M.~Serrani, ``{Associativity of celestial OPE, higher spins and self-duality},'' {\em JHEP} {\bf 04} (2026) 047, \href{http://arXiv.org/abs/2508.16804}{{\tt 2508.16804}}.

\bibitem{Ren:2022sws}
L.~Ren, M.~Spradlin, A.~Yelleshpur~Srikant, and A.~Volovich, ``{On effective field theories with celestial duals},'' {\em JHEP} {\bf 08} (2022) 251, \href{http://arXiv.org/abs/2206.08322}{{\tt 2206.08322}}.

\bibitem{Monteiro:2011pc}
R.~Monteiro and D.~O'Connell, ``{The Kinematic Algebra From the Self-Dual Sector},'' {\em JHEP} {\bf 07} (2011) 007, \href{http://arXiv.org/abs/1105.2565}{{\tt 1105.2565}}.

\bibitem{Devchand:1996gv}
C.~Devchand and V.~Ogievetsky, ``{Interacting fields of arbitrary spin and $N >4$ supersymmetric selfdual Yang-Mills equations},'' {\em Nucl. Phys. B} {\bf 481} (1996) 188--214, \href{http://arXiv.org/abs/hep-th/9606027}{{\tt hep-th/9606027}}.

\bibitem{Sharapov:2022awp}
A.~Sharapov and E.~Skvortsov, ``{Chiral higher spin gravity in (A)dS4 and secrets of Chern{\textendash}Simons matter theories},'' {\em Nucl. Phys. B} {\bf 985} (2022) 115982, \href{http://arXiv.org/abs/2205.15293}{{\tt 2205.15293}}.

\bibitem{Aharony:2024nqs}
O.~Aharony, R.~R. Kalloor, and T.~Kukolj, ``{A chiral limit for Chern-Simons-matter theories},'' {\em JHEP} {\bf 10} (2024) 051, \href{http://arXiv.org/abs/2405.01647}{{\tt 2405.01647}}.

\bibitem{Jain:2024bza}
S.~Jain, D.~K. S, and E.~Skvortsov, ``{Hidden sectors of Chern-Simons matter theories and exact holography},'' {\em Phys. Rev. D} {\bf 111} (2025), no.~10, 106017, \href{http://arXiv.org/abs/2405.00773}{{\tt 2405.00773}}.

\bibitem{Armstrong:2020woi}
C.~Armstrong, A.~E. Lipstein, and J.~Mei, ``{Color/kinematics duality in AdS$_{4}$},'' {\em JHEP} {\bf 02} (2021) 194, \href{http://arXiv.org/abs/2012.02059}{{\tt 2012.02059}}.

\bibitem{Chowdhury:2024dcy}
C.~Chowdhury, G.~Doran, A.~Lipstein, R.~Monteiro, S.~Nagy, and K.~Singh, ``{Light-cone actions and correlators of self-dual theories in AdS$_{4}$},'' {\em JHEP} {\bf 01} (2025) 172, \href{http://arXiv.org/abs/2411.04172}{{\tt 2411.04172}}.

\bibitem{Skvortsov:2026gtq}
E.~Skvortsov and R.~Van~Dongen, ``{Dirichlet, Neumann, mixed and self-dual holography: (self-dual) Yang-Mills theory},'' {\em JHEP} {\bf 09} (2026) 282, \href{http://arXiv.org/abs/2602.21658}{{\tt 2602.21658}}.

\bibitem{Guevara:2026qzd}
A.~Guevara, A.~Lupsasca, D.~Skinner, A.~Strominger, and K.~Weil, ``{Single-minus gluon tree amplitudes are nonzero},'' \href{http://arXiv.org/abs/2602.12176}{{\tt 2602.12176}}.

\bibitem{Guevara:2026qwa}
A.~Guevara, A.~Lupsasca, D.~Skinner, A.~Strominger, and K.~Weil, ``{Single-minus graviton tree amplitudes are nonzero},'' \href{http://arXiv.org/abs/2603.04330}{{\tt 2603.04330}}.

\bibitem{DelDuca:1999iql}
V.~Del~Duca, A.~Frizzo, and F.~Maltoni, ``{Factorization of tree QCD amplitudes in the high-energy limit and in the collinear limit},'' {\em Nucl. Phys. B} {\bf 568} (2000) 211--262, \href{http://arXiv.org/abs/hep-ph/9909464}{{\tt hep-ph/9909464}}.

\bibitem{DelDuca:1999rs}
V.~Del~Duca, L.~J. Dixon, and F.~Maltoni, ``{New color decompositions for gauge amplitudes at tree and loop level},'' {\em Nucl. Phys. B} {\bf 571} (2000) 51--70, \href{http://arXiv.org/abs/hep-ph/9910563}{{\tt hep-ph/9910563}}.

\bibitem{Monteiro:2022lwm}
R.~Monteiro, ``{Celestial chiral algebras, colour-kinematics duality and integrability},'' {\em JHEP} {\bf 01} (2023) 092, \href{http://arXiv.org/abs/2208.11179}{{\tt 2208.11179}}.

\bibitem{Klebanov:2002ja}
I.~R. Klebanov and A.~M. Polyakov, ``{AdS dual of the critical $O(N)$ vector model},'' {\em Phys. Lett.} {\bf B550} (2002) 213--219,
\href{http://arXiv.org/abs/hep-th/0210114}{{\tt hep-th/0210114}}.

\bibitem{Sezgin:2002rt}
E.~Sezgin and P.~Sundell, ``{Massless higher spins and holography},'' {\em Nucl.Phys.} {\bf B644} (2002) 303--370,
\href{http://arXiv.org/abs/hep-th/0205131}{{\tt hep-th/0205131}}.

\bibitem{Leigh:2003gk}
R.~G. Leigh and A.~C. Petkou, ``{Holography of the N=1 higher spin theory on AdS(4)},'' {\em JHEP} {\bf 06} (2003) 011, \href{http://arXiv.org/abs/hep-th/0304217}{{\tt hep-th/0304217}}.

\bibitem{Ponomarev:2022qkx}
D.~Ponomarev, ``{Chiral higher-spin holography in flat space: the Flato-Fronsdal theorem and lower-point functions},'' {\em JHEP} {\bf 01} (2023) 048, \href{http://arXiv.org/abs/2210.04036}{{\tt 2210.04036}}.

\bibitem{Ponomarev:2022ryp}
D.~Ponomarev, ``{Towards higher-spin holography in flat space},'' {\em JHEP} {\bf 01} (2023) 084, \href{http://arXiv.org/abs/2210.04035}{{\tt 2210.04035}}.

\bibitem{Chalmers:1996rq}
G.~Chalmers and W.~Siegel, ``{The Selfdual sector of QCD amplitudes},'' {\em Phys. Rev.} {\bf D54} (1996) 7628--7633,
\href{http://arXiv.org/abs/hep-th/9606061}{{\tt hep-th/9606061}}.

\bibitem{Berends:1987me}
F.~A. Berends and W.~T. Giele, ``{Recursive Calculations for Processes with n Gluons},'' {\em Nucl. Phys. B} {\bf 306} (1988) 759--808.

\bibitem{Penrose:1965am}
R.~Penrose, ``{Zero rest mass fields including gravitation: Asymptotic behavior},'' {\em Proc. Roy. Soc. Lond.} {\bf A284} (1965)
159.

\bibitem{Hughston:1979tq}
L.~P. Hughston, R.~S. Ward, M.~G. Eastwood, M.~L. Ginsberg, A.~P. Hodges, S.~A. Huggett, T.~R. Hurd, R.~O. Jozsa, R.~Penrose, A.~Popovich, {\em et al.}, eds., {\em {Advances in twistor theory}}.
\newblock
1979.
\newblock

\bibitem{Eastwood:1981jy}
M.~G. Eastwood, R.~Penrose, and R.~O. Wells, ``{Cohomology and Massless Fields},'' {\em Commun. Math. Phys.} {\bf 78} (1981)
305--351.

\bibitem{Woodhouse:1985id}
N.~M.~J. Woodhouse, ``{Real methods in twistor theory},'' {\em Class. Quant. Grav.} {\bf 2} (1985)
257--291.

\bibitem{Guarini:2026vds}
R.~Guarini, ``{On amplitudes in chiral higher spin gravity},'' {\em JHEP} {\bf 06} (2026) 034, \href{http://arXiv.org/abs/2603.06044}{{\tt 2603.06044}}.

\bibitem{Tran:2022amg}
T.~Tran, ``{Constraining higher-spin S-matrices},'' {\em JHEP} {\bf 02} (2023) 001, \href{http://arXiv.org/abs/2212.02540}{{\tt 2212.02540}}.

\bibitem{Bengtsson:1983pd}
A.~K.~H. Bengtsson, I.~Bengtsson, and L.~Brink, ``{Cubic interaction terms for arbitrary spin},'' {\em Nucl. Phys.} {\bf B227} (1983)
31.

\bibitem{Bengtsson:1986kh}
A.~K.~H. Bengtsson, I.~Bengtsson, and N.~Linden, ``{Interacting Higher Spin Gauge Fields on the Light Front},'' {\em Class. Quant. Grav.} {\bf 4} (1987)
1333.

\bibitem{Ponomarev:2022vjb}
D.~Ponomarev, ``{Basic Introduction to Higher-Spin Theories},'' {\em Int. J. Theor. Phys.} {\bf 62} (2023), no.~7, 146, \href{http://arXiv.org/abs/2206.15385}{{\tt 2206.15385}}.

\bibitem{penroserindler}
R.~Penrose and W.~Rindler, {\em Spinors and Space-Time}, vol.~1 of {\em Cambridge Monographs on Mathematical Physics}.
\newblock Cambridge University Press, 1984.

\bibitem{Benincasa:2011pg}
P.~Benincasa and E.~Conde, ``{Exploring the S-Matrix of Massless Particles},'' {\em Phys. Rev. D} {\bf 86} (2012) 025007, \href{http://arXiv.org/abs/1108.3078}{{\tt 1108.3078}}.

\bibitem{Bern:2010ue}
Z.~Bern, J.~J.~M. Carrasco, and H.~Johansson, ``{Perturbative Quantum Gravity as a Double Copy of Gauge Theory},'' {\em Phys. Rev. Lett.} {\bf 105} (2010) 061602, \href{http://arXiv.org/abs/1004.0476}{{\tt 1004.0476}}.

\bibitem{Strachan:1992em}
I.~A.~B. Strachan, ``{The Moyal algebra and integrable deformations of the selfdual Einstein equations},'' {\em Phys. Lett. B} {\bf 283} (1992) 63--66.

\bibitem{Bu:2022iak}
W.~Bu, S.~Heuveline, and D.~Skinner, ``{Moyal deformations, W$_{1+\infty}$ and celestial holography},'' {\em JHEP} {\bf 12} (2022) 011, \href{http://arXiv.org/abs/2208.13750}{{\tt 2208.13750}}.

\bibitem{Bittleston:2023bzp}
R.~Bittleston, S.~Heuveline, and D.~Skinner, ``{The celestial chiral algebra of self-dual gravity on Eguchi-Hanson space},'' {\em JHEP} {\bf 09} (2023) 008, \href{http://arXiv.org/abs/2305.09451}{{\tt 2305.09451}}.

\bibitem{Konstein:1989ij}
S.~E. Konstein and M.~A. Vasiliev, ``Extended higher spin superalgebras and their massless representations,'' {\em Nucl. Phys.} {\bf B331} (1990)
475--499.

\bibitem{Tran:2021ukl}
T.~Tran, ``{Twistor constructions for higher-spin extensions of (self-dual) Yang-Mills},'' {\em JHEP} {\bf 11} (2021) 117, \href{http://arXiv.org/abs/2107.04500}{{\tt 2107.04500}}.

\bibitem{Herfray:2022prf}
Y.~Herfray, K.~Krasnov, and E.~Skvortsov, ``{Higher-spin self-dual Yang-Mills and gravity from the twistor space},'' {\em JHEP} {\bf 01} (2023) 158, \href{http://arXiv.org/abs/2210.06209}{{\tt 2210.06209}}.

\bibitem{Tran:2022tft}
T.~Tran, ``{Toward a twistor action for chiral higher-spin gravity},'' {\em Phys. Rev. D} {\bf 107} (2023), no.~4, 046015, \href{http://arXiv.org/abs/2209.00925}{{\tt 2209.00925}}.

\bibitem{Mason:2025pbz}
L.~Mason and A.~Sharma, ``{Chiral higher-spin theories from twistor space},'' \href{http://arXiv.org/abs/2505.09419}{{\tt 2505.09419}}.

\bibitem{Kleiss:1988ne}
R.~Kleiss and H.~Kuijf, ``{Multi - Gluon Cross-sections and Five Jet Production at Hadron Colliders},'' {\em Nucl. Phys. B} {\bf 312} (1989) 616--644.

\bibitem{Johansson:2015oia}
H.~Johansson and A.~Ochirov, ``{Color-Kinematics Duality for QCD Amplitudes},'' {\em JHEP} {\bf 01} (2016) 170, \href{http://arXiv.org/abs/1507.00332}{{\tt 1507.00332}}.

\bibitem{Bern:2011ia}
Z.~Bern and T.~Dennen, ``{A Color Dual Form for Gauge-Theory Amplitudes},'' {\em Phys. Rev. Lett.} {\bf 107} (2011) 081601, \href{http://arXiv.org/abs/1103.0312}{{\tt 1103.0312}}.

\bibitem{Fu:2018hpu}
C.-H. Fu, P.~Vanhove, and Y.~Wang, ``{A Vertex Operator Algebra Construction of the Colour-Kinematics Dual numerator},'' {\em JHEP} {\bf 09} (2018) 141, \href{http://arXiv.org/abs/1806.09584}{{\tt 1806.09584}}.

\bibitem{Bjerrum-Bohr:2012kaa}
N.~E.~J. Bjerrum-Bohr, P.~H. Damgaard, R.~Monteiro, and D.~O'Connell, ``{Algebras for Amplitudes},'' {\em JHEP} {\bf 06} (2012) 061, \href{http://arXiv.org/abs/1203.0944}{{\tt 1203.0944}}.

\bibitem{Bardeen:1995gk}
W.~A. Bardeen, ``{Selfdual Yang-Mills theory, integrability and multiparton amplitudes},'' {\em Prog. Theor. Phys. Suppl.} {\bf 123} (1996) 1--8.

\bibitem{Krasnov:2016emc}
K.~Krasnov, ``{Self-Dual Gravity},'' {\em Class. Quant. Grav.} {\bf 34} (2017), no.~9, 095001, \href{http://arXiv.org/abs/1610.01457}{{\tt 1610.01457}}.

\bibitem{Hasuwannakit:2025agr}
C.~Hasuwannakit and K.~Krasnov, ``{Gravity MHV amplitudes via Berends-Giele currents},'' {\em JHEP} {\bf 11} (2025) 156, \href{http://arXiv.org/abs/2507.13943}{{\tt 2507.13943}}.

\bibitem{Witten:2003nn}
E.~Witten, ``{Perturbative gauge theory as a string theory in twistor space},'' {\em Commun. Math. Phys.} {\bf 252} (2004) 189--258, \href{http://arXiv.org/abs/hep-th/0312171}{{\tt hep-th/0312171}}.

\bibitem{Roiban:2004yf}
R.~Roiban, M.~Spradlin, and A.~Volovich, ``{On the tree level S matrix of Yang-Mills theory},'' {\em Phys. Rev. D} {\bf 70} (2004) 026009, \href{http://arXiv.org/abs/hep-th/0403190}{{\tt hep-th/0403190}}.

\bibitem{Boels:2013bi}
R.~H. Boels, R.~S. Isermann, R.~Monteiro, and D.~O'Connell, ``{Colour-Kinematics Duality for One-Loop Rational Amplitudes},'' {\em JHEP} {\bf 04} (2013) 107, \href{http://arXiv.org/abs/1301.4165}{{\tt 1301.4165}}.

\bibitem{Ponomarev:2016cwi}
D.~Ponomarev, ``{Off-Shell Spinor-Helicity Amplitudes from Light-Cone Deformation Procedure},'' {\em JHEP} {\bf 12} (2016) 117,
\href{http://arXiv.org/abs/1611.00361}{{\tt 1611.00361}}.

\bibitem{Cachazo:2004kj}
F.~Cachazo, P.~Svrcek, and E.~Witten, ``{MHV vertices and tree amplitudes in gauge theory},'' {\em JHEP} {\bf 09} (2004) 006, \href{http://arXiv.org/abs/hep-th/0403047}{{\tt hep-th/0403047}}.

\bibitem{Gorsky:2005sf}
A.~Gorsky and A.~Rosly, ``{From Yang-Mills Lagrangian to MHV diagrams},'' {\em JHEP} {\bf 01} (2006) 101, \href{http://arXiv.org/abs/hep-th/0510111}{{\tt hep-th/0510111}}.

\bibitem{Mansfield:2005yd}
P.~Mansfield, ``{The Lagrangian origin of MHV rules},'' {\em JHEP} {\bf 03} (2006) 037, \href{http://arXiv.org/abs/hep-th/0511264}{{\tt hep-th/0511264}}.

\bibitem{Skvortsov:2022syz}
E.~Skvortsov and R.~Van~Dongen, ``{Minimal models of field theories: Chiral higher spin gravity},'' {\em Phys. Rev. D} {\bf 106} (2022), no.~4, 045006, \href{http://arXiv.org/abs/2204.10285}{{\tt 2204.10285}}.

\bibitem{Sharapov:2022faa}
A.~Sharapov, E.~Skvortsov, A.~Sukhanov, and R.~Van~Dongen, ``{Minimal model of Chiral Higher Spin Gravity},'' {\em JHEP} {\bf 09} (2022) 134, \href{http://arXiv.org/abs/2205.07794}{{\tt 2205.07794}}. [Erratum: JHEP 02, 183 (2023)].

\bibitem{LepowskyLi2004}
J.~Lepowsky and H.~Li, {\em Introduction to Vertex Operator Algebras and Their Representations}, vol.~227 of {\em Progress in Mathematics}.
\newblock Birkh{\"a}user Boston, Boston, MA, 2004.

\bibitem{nozaradan2008introductionvertexalgebras}
C.~Nozaradan, ``Introduction to vertex algebras,'' 2008.

\bibitem{Kawai:1985xq}
H.~Kawai, D.~C. Lewellen, and S.~H.~H. Tye, ``{A Relation Between Tree Amplitudes of Closed and Open Strings},'' {\em Nucl. Phys. B} {\bf 269} (1986) 1--23.

\bibitem{Cachazo:2013iea}
F.~Cachazo, S.~He, and E.~Y. Yuan, ``{Scattering of Massless Particles: Scalars, Gluons and Gravitons},'' {\em JHEP} {\bf 07} (2014) 033, \href{http://arXiv.org/abs/1309.0885}{{\tt 1309.0885}}.

\bibitem{Brandhuber:2026njb}
A.~Brandhuber, P.~Pichini, G.~Travaglini, and C.~Wen, ``{{\ensuremath{{N}}} = 4 single-minus superamplitudes and dual superconformal symmetry},'' {\em JHEP} {\bf 10} (2026) 171, \href{http://arXiv.org/abs/2603.26609}{{\tt 2603.26609}}.

\bibitem{Sharapov:2022wpz}
A.~Sharapov, E.~Skvortsov, and R.~Van~Dongen, ``{Chiral higher spin gravity and convex geometry},'' {\em SciPost Phys.} {\bf 14} (2023), no.~6, 162, \href{http://arXiv.org/abs/2209.01796}{{\tt 2209.01796}}.

\bibitem{Sezgin:2011hq}
E.~Sezgin and P.~Sundell, ``{Geometry and Observables in Vasiliev's Higher Spin Gravity},'' {\em JHEP} {\bf 07} (2012) 121,
\href{http://arXiv.org/abs/1103.2360}{{\tt 1103.2360}}.

\bibitem{Colombo:2012jx}
N.~Colombo and P.~Sundell, ``{Higher Spin Gravity Amplitudes From Zero-form Charges},''
\href{http://arXiv.org/abs/1208.3880}{{\tt 1208.3880}}.

\bibitem{Didenko:2012tv}
V.~Didenko and E.~Skvortsov, ``{Exact higher-spin symmetry in CFT: all correlators in unbroken Vasiliev theory},'' {\em JHEP} {\bf 1304} (2013) 158,
\href{http://arXiv.org/abs/1210.7963}{{\tt 1210.7963}}.

\bibitem{Bonezzi:2017vha}
R.~Bonezzi, N.~Boulanger, D.~De~Filippi, and P.~Sundell, ``{Noncommutative Wilson lines in higher-spin theory and correlation functions of conserved currents for free conformal fields},'' {\em J. Phys.} {\bf A50} (2017), no.~47, 475401,
\href{http://arXiv.org/abs/1705.03928}{{\tt 1705.03928}}.

\bibitem{Strominger:2017zoo}
A.~Strominger, {\em {Lectures on the Infrared Structure of Gravity and Gauge Theory}}.
\newblock Princeton University Press, 2018.
\newblock \href{http://arXiv.org/abs/1703.05448}{{\tt 1703.05448}}.

\bibitem{Maldacena:2011nz}
J.~M. Maldacena and G.~L. Pimentel, ``{On graviton non-Gaussianities during inflation},'' {\em JHEP} {\bf 09} (2011) 045, \href{http://arXiv.org/abs/1104.2846}{{\tt 1104.2846}}.

\end{thebibliography}
\end{document}